\documentclass{elsarticle}

\usepackage[margin=2.5cm]{geometry}
\usepackage[utf8]{inputenc}
\usepackage{graphics}
\usepackage{graphicx}
\usepackage{multicol}
\usepackage{color}
\usepackage{tabularx}
\usepackage{longtable}
\usepackage{graphics,epsfig,aas_macros,amsmath}

\begin{document}
	
	\title{Thermonuclear X-ray bursts detected in Cyg X-2 using AstroSat/LAXPC}
	
	\author[Devasia et al.]
	{Jincy Devasia$^1$, Gayathri Raman$^{2,3}$ and Biswajit Paul$^4$  \\ 
		$^1$Henry Baker College, Melukavu, Kottayam-686 652, Kerala, India\\
		$^2$International Center for Theoretical Sciences (ICTS-TIFR), Heseraghatta, Bengaluru-560089, India \\
		$^3$Indian Institute of Technology (IIT) Bombay, Main Gate Rd, IIT Area, Powai, Mumbai 400076, India\\  
		$^4$Department of Astronomy and Astrophysics, Raman Research Institute, Sadashivanagar, Bangalore-560080, India}
	
	\date{}
	
	
	\begin{frontmatter}
	\begin{abstract}
		We report the detection of 5 Type-1 thermonuclear X-ray bursts and one burst-like event in the neutron star LMXB source, Cyg X-2 using X-ray data obtained with the Large Area X-ray Proportional Counter (LAXPC) instrument on board AstroSat. We carry out an energy resolved burst profile analysis as well as time resolved spectral analysis for each of the bursts and characterize their properties. All bursts are weak with burst peak-to-persistent intensity ratios $<$ 3, decay times $\sim$2 s, and with fluences $\sim$1$\times$10$^{-8}$ ergs/cm$^2$, indicating that the observed bursts are Helium fuelled flashes. An evolution of the blackbody temperature and radius is also observed during each burst. We carry out a search for Burst Oscillations (BO) and derive upper limits to the rms fractional amplitude for BO (for all the bursts) to be $\sim$1\%. We also carried out search for Quasi Periodic Oscillations (QPOs) in the power density spectra and we obtain upper limits to the fractional rms amplitude as $\sim$3.4\% at frequencies close to $\sim$5.6 Hz.  We further carry out spectral and timing analysis of the non-burst persistent emission along with a study of the hardness-intensity and colour-colour diagrams. Using results from our analysis we infer that during this observation in 2016, Cyg X-2 can be characterized as being in the early Flaring Branch (FB) with a puffed up accretion disk and a clumpy coronal structure while undergoing medium-to-high levels of accretion.

	\end{abstract}
	
	\begin{keyword}
		X-rays: binaries, stars: neutron, accretion, accretion discs
		\end{keyword}
	
	\end{frontmatter}
	
	\section{Introduction}
	Cyg X-2 is a neutron star Low Mass X-ray Binary (LMXB) discovered in 1965 \citep{Bowyer1965}. It is a persistent X-ray binary that has been observed to be steadily accreting near its Eddington limit \citep{KG1984,Smale1998}. The companion is an evolved, late-type star with mass ranging between 0.4 and 0.7 M$\odot$, and a spectral type described between A5 to F2 \citep{CCH1979}. The average long-term flux in the 2.5-25 keV band is 11$\times$10$^{-9}$ ergs/cm$^{2}$/s \citep{Galloway2008}. The binary orbital period for this source is 9.844$\pm$0.0003 d \citep{CCH1979}. Assuming a binary inclination of $\sim$ 62.5$\pm$4$^{\circ}$, the implied mass of the neutron star is 1.71$\pm$0.21 M$\odot$ \citep{Casares2010}. 
	
	Since Cyg X-2 follows a Z-shaped spectral pattern in the X-ray colour-colour diagram (CCD), it is classified as a Z source \citep{HV1989}. Cyg like sources (eg. Cyg X-2, GX 1+5) move significantly along all branches of the Z-track - the Horizontal Branch (HB), the Normal Branch (NB), and the Flaring Branch (FB).  The earliest understanding has been that along the Z track, the mass accretion rate evolves monotonically in the direction HB-NB-FB \citep{Pri1986,Hasinger1990}. This has been debated in more recent literature (see  \citealt{Church2006,Lin2009}). \citet{Homan2010} further suggested that the position along the color-color tracks of Z sources is not determined by the instantaneous mass accretion rate. A detailed review of the nature of Cyg-like Z-track sources with regards to their transitions between the different branches, their timing and spectral properties has been presented by \citet{Church2010}. During the 2015 NuSTAR observations, this source exhibited dips and strong variability in 3-79 keV X-ray light curve \citep{Mondal2018}. The source remained in the NB of the Z-track during non-dipping epochs, while it transitioned to the FB during dips. Cyg X-2 is also known for exhibiting secular variations within the Z track \citep{Wijnands1997}. 
	
	Broad band spectral studies of Cyg X-2 have been carried out using Beppo-SAX and NuSTAR \citep{DiSalvo2002,Mondal2018}. The broadband continuum has traditionally been described using the `Eastern' model comprising of a disk blackbody component along with a thermal compotonization component \citep{DiSalvo2002}. Some studies have additionally utilized a reflection model in the continuum (for example, \citealt{Mondal2018}) and some have invoked the presence of an Accretion Disk Corona (ADC) to explain the flux states as a consequence of geometrical and optical disk thickness \citep{Schulz2009, Church2010}. Several spectral lines have also been reported like the Fe complex in the 6-7 keV range, H-like Ne, Mg, Si lines \citep{Shaposhnikov2009,Mondal2018}. \citet{Piraino2001} also observed a hard X-ray tail in the source spectra which they attribute to high energy non-thermal emission.
	
	Cyg X-2 has also displayed a number of Type-1 thermonuclear X-ray bursts and burst-like events in the past which established the neutron star nature of the primary compact object \citep{KG1984,Smale1998}. Using data obtained with the EXOSAT mission, \citet{Kuulkers1995} reported detections of around 9 burst-like events, which were described as having profiles close to Type-1 X-ray bursts. \citet{Wijnands1997} also reported a similar detection of 5 burst-like events using Ginga data. The first confirmed report of a burst was by \citet{Smale1998} and it was a Photospheric Radius Expansion (PRE) burst. Powerful bursting can result in radial atmosphere expansion, where after the burst peak, the luminosity falls below the Eddington limit as the atmosphere quickly contracts back to the neutron star surface. This PRE burst was used to estimate the source distance to be $\sim$11 kpc, while optical observations indicated a distance of about 7.2$\pm$1.1 kpc \citep{O-K1999}. Further analysis of RXTE data sets confirmed the presence of several other PRE thermonuclear bursts and burst-like events \citep{Tit2002}. The bursts are typically observed with durations of about 5 s. \citet{Kuulkers1995} and \citet{Smale1998} placed an upper limit of 10\% (1-256 Hz) and 2\% (200-600 Hz) on burst oscillation (BO) amplitudes  using EXOSAT and RXTE, respectively.
	
	Cyg X-2 has previously shown a number of Quasi Periodic Oscillation (QPO) features in its power density spectra. Twin kHz QPOs around 500 and 860 Hz and also the single highest QPO at 1007 Hz were discovered using RXTE when it was in the HB \citep{Wijnands1998}. The source also exhibited 18-50 Hz QPOs in the HB,  5-6 Hz slow QPOs and 50-60 Hz QPOs in the NB or NB/FB vertex \citep{Elsner1986,Hasinger1986,Hasinger1990,Wijnands1997,Wijnands1998}. However, there have not been many previous reports of QPOs being detected during the FB alone (see \citealt{Kuul1995} for a specialized case). \citet{Kuul1995} report a $\sim$26 Hz QPO only during the light curve `dip' corresponding to the FB, and not during the other FB epochs. From the frequency difference in the twin kHz QPO peaks, the inferred spin of the neutron star was 346$\pm$29 Hz \citep{Wijnands1998}, although it has been alternatively noted that the kHz QPO peak difference frequency is independent of the neutron star spin frequency \citep{Mendez_Belloni2007}.

	In this work we present the results of timing and spectral analysis of AstroSat/LAXPC observations of Cyg X-2 carried out in 2016 and report the detection of 5 type 1 thermonuclear X-ray bursts and one burst-like event. The paper is organised as follows. In Section 2 we describe the details of the AstroSat observations followed by timing and spectral analysis methods and results in Section 3.  Using the results obtained from energy resolved burst light curve analysis, time resolved burst spectral analysis, power spectra and colour-intensity analysis, we construct a coherent picture of the source state and discuss the implications in Section 4.

	\section{Observations}
	
	AstroSat is India's first dedicated multi-wavelength astronomy satellite mission, launched successfully in 2015. The satellite comprises of five payloads that are sensitive from the UV/optical all the way up to the hard X-ray energy bands. The LAXPC is a broad band timing instrument with a 1$^{\circ}$ $\times$ 1$^{\circ}$ field of view. It has a timing resolution of $\sim$10$\mu$s, making it an ideal instrument to probe fast processes in neutron star systems such as milli-second pulsations in rapidly rotating objects, QPOs and Burst Oscillations (BO). The LAXPC is sensitive in the wide energy band of 3.0 - 80 keV and has three co-aligned but independent proportional counter detector units LXP10, LXP20 and LXP30. Details of the LAXPC instrument and calibration are presented in \citet{Yadav2017}. Cyg X-2 was observed using LAXPC on the 28th and 29th of February, 2016 as part of the Guaranteed Time (GT) observation phase, details of the observations are given in Table \ref{tab:obs}. \\

	\begin{table*}
		
		\begin{tabular}{|c|c|c|c|c|}
			\hline
			Obs ID & Orbits & Instrument & Obs date & Orbit combined Exposure (ks)  \\
			\hline
			348 & 2272-2296 & LAXPC & 28th, 29th February, 2016  & 150 \\
			\hline
		\end{tabular}
		\caption{A list of all the AstroSat observations of Cyg X-2 using LAXPC.}
		\label{tab:obs}
	\end{table*}

	\section{Data Analysis and Results}

\begin{figure}
		\centering
		\includegraphics[scale=0.45,angle=0]{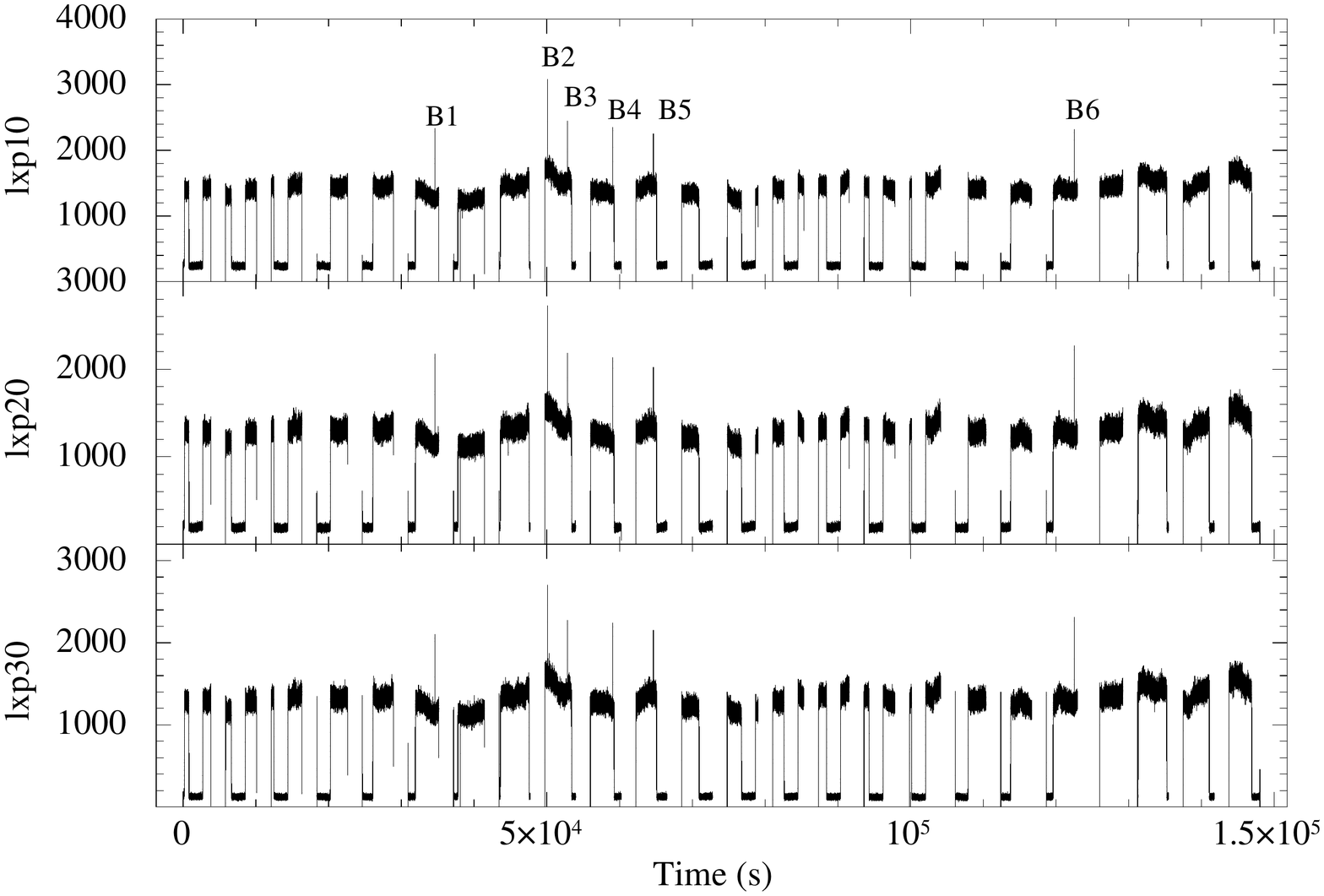}
		\caption{The raw light curve of CygX-2 from lxp10, lxp20 and lxp30 with a binning of 1 s. There are 6 thermonuclear bursts/burst-like events detected. }
		\label{fig:lc}
	\end{figure}

	The LAXPC observations of Cyg X-2 were carried out in the Event Analysis (EA) mode with an exposure time of 150 ks. The raw Level 1 data files were reduced using the LAXPC data analysis pipeline version 3.1\footnote{Data anlysis software was obtained from http://astrosat-ssc.iucaa.in/?q=laxpcData}.  A good time interval window was applied during the processing in order to eliminate time intervals corresponding to the Earth occultation periods and SAA passage. Standard elevation angle cuts were also applied as part of the Level 1 processing. The resultant Level 2 event files were then used to extract the 0.1 ms binned light curves using the task \textit{lcphaextrct} from all the 7 detector layers. All single and double events were also included. The background subtracted 1 s binned light curves from all the LAXPC detectors are shown in Figure \ref{fig:lc}. 

\begin{figure}
		\centering
		\includegraphics[scale=0.4,angle=0]{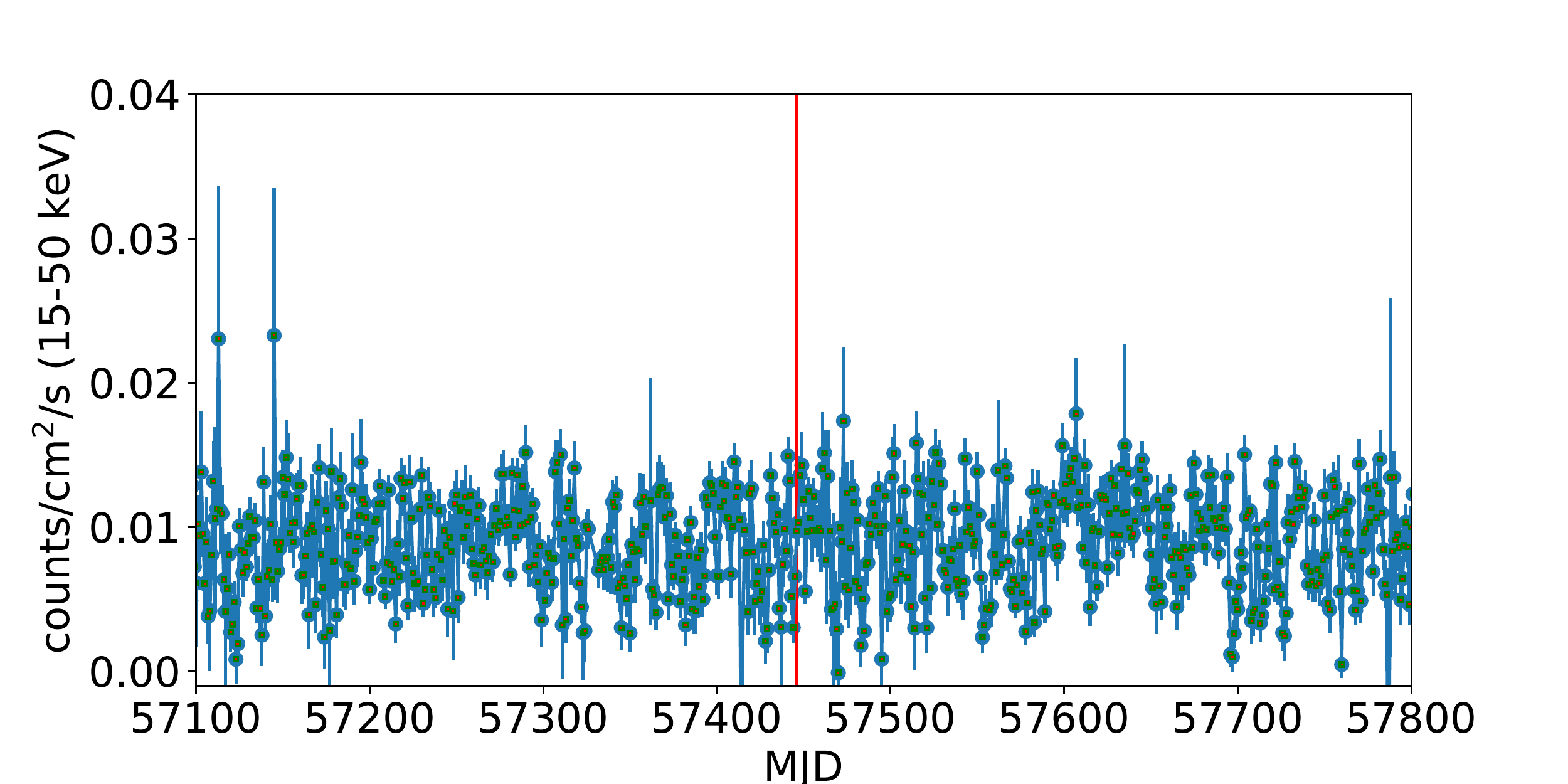}
		\caption{The 15-50 keV BAT light curve of Cyg X-2 indicating the date of LAXPC observation. }
		\label{fig:lc2}
	\end{figure}

	In order to determine the overall intensity of the source, we have marked the position of the AstroSat observations on the long term 15-50 keV Swift-BAT light curve (see Figure \ref{fig:lc2}).

	\subsection{Type 1 X-ray bursts}
	
	
	\subsubsection{Burst timing analysis}
	
	Light curves from the LAXPC detectors detected simultaneous presence of a total of six X-ray burst-like peaks. The light curves from the three detectors were co-added using the ftools task $\lq$\textit{lcmath}' for further analysis. The fast linear rise and exponential decay profile (FRED), characteristic of a Type-1 burst is discernible in most of the detected bursts. The burst characteristics are elaborated in Table \ref{tab:burs} where each of the bursts are numbered sequentially from B1 to B6. Particularly, B5 exhibits a sub structure with two peaks, distinguishing it from the remaining bursts. We call this a burst-like event, much similar to the burst-like events observed in Cyg X-2 previously (see \citealt{Wijnands1997,Kuulkers1995}). B1 and B2 are separated by about 4 hours whereas the following 4 bursts i.e., B2, B3, B4 and B5 appear in rapid succession separated by an hour. The last burst, B6 is again spaced out more and occurs $\sim$13 hours after B5. We note that we could have missed possible bursts during the data gaps.
	
\begin{figure}
		 \begin{center}
		\begin{minipage}{1.0\linewidth}
			\includegraphics[scale=0.35,trim={5cm 5cm 8cm 6cm},clip=true,angle=0]{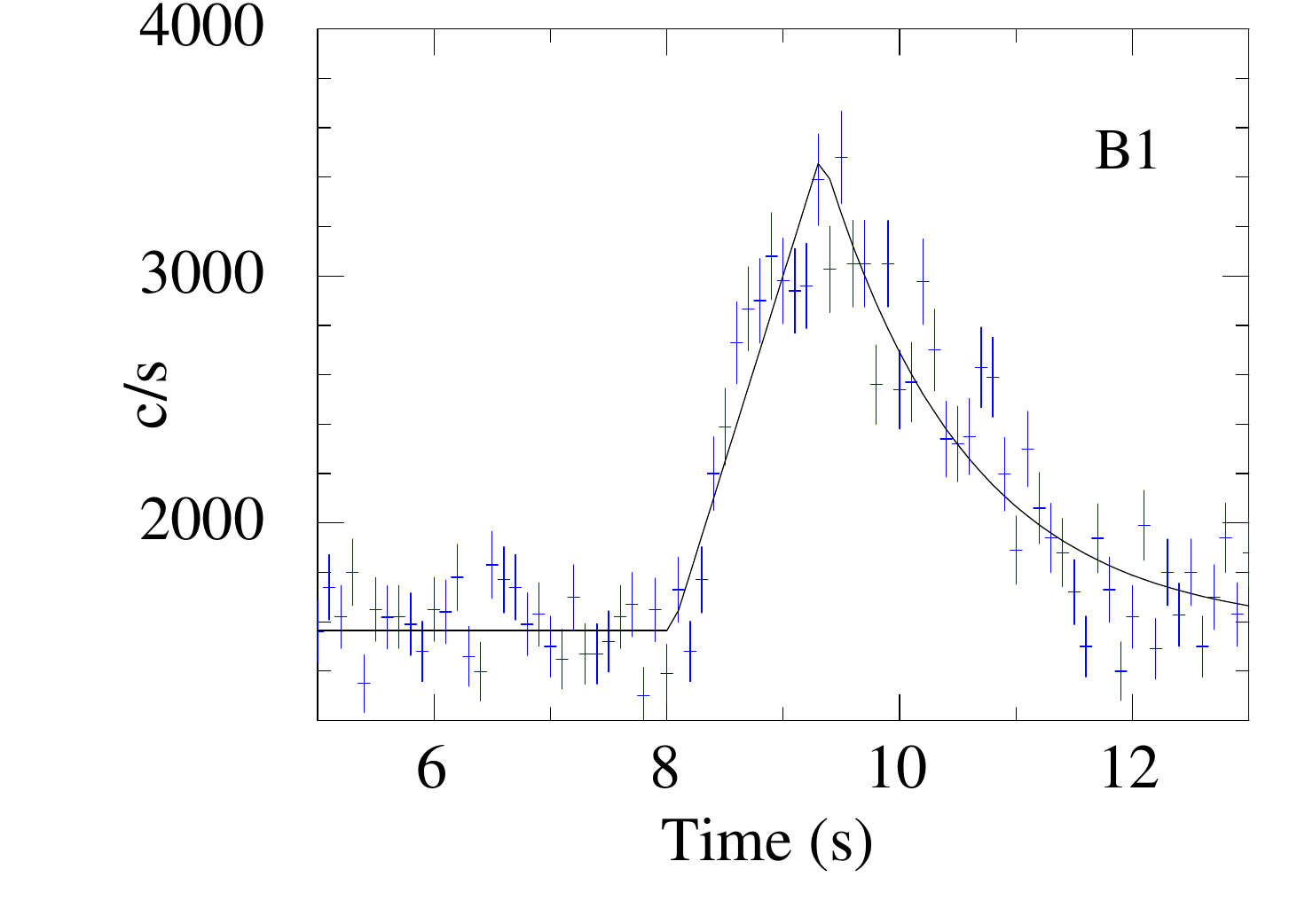}
			\includegraphics[scale=0.35,trim={5cm 5cm 8cm 6cm},clip=true,angle=0]{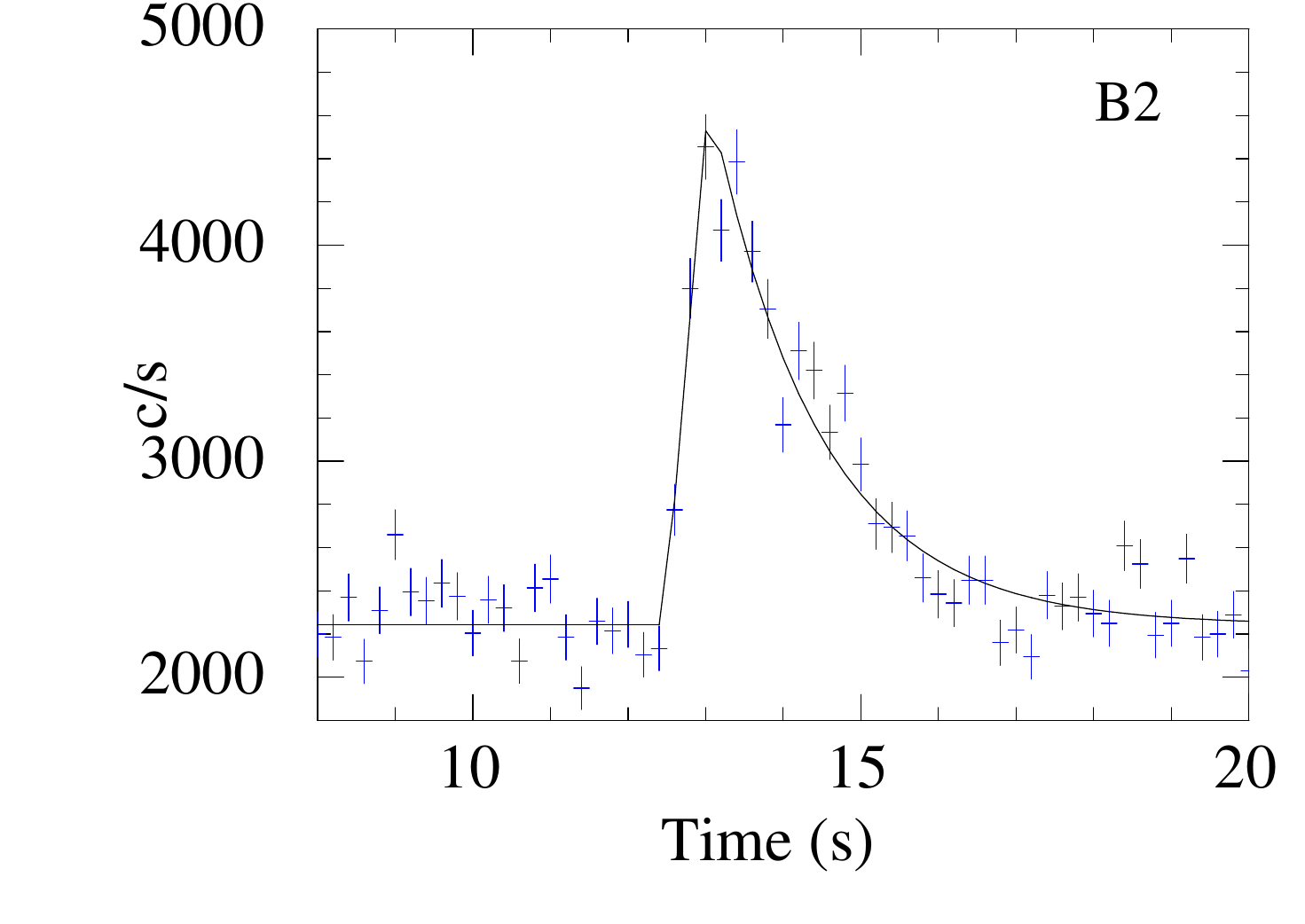}
			\includegraphics[scale=0.35,trim={5cm 5cm 8cm 6cm},clip=true,angle=0]{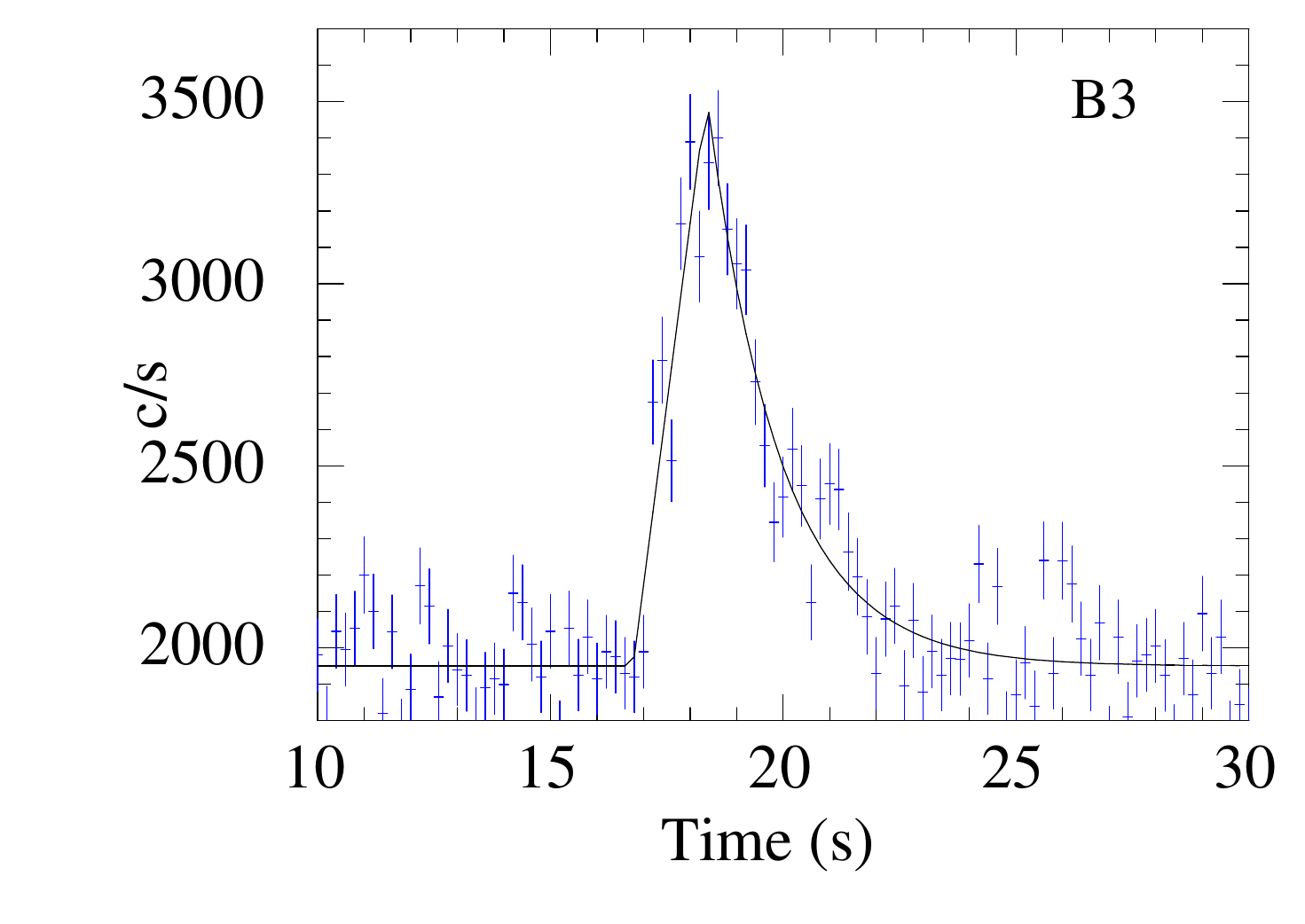}
		\end{minipage}     
		
		\begin{minipage}{1.0\linewidth}   

			\includegraphics[scale=0.35,trim={5cm 5cm 8cm 6cm},clip=true,angle=0]{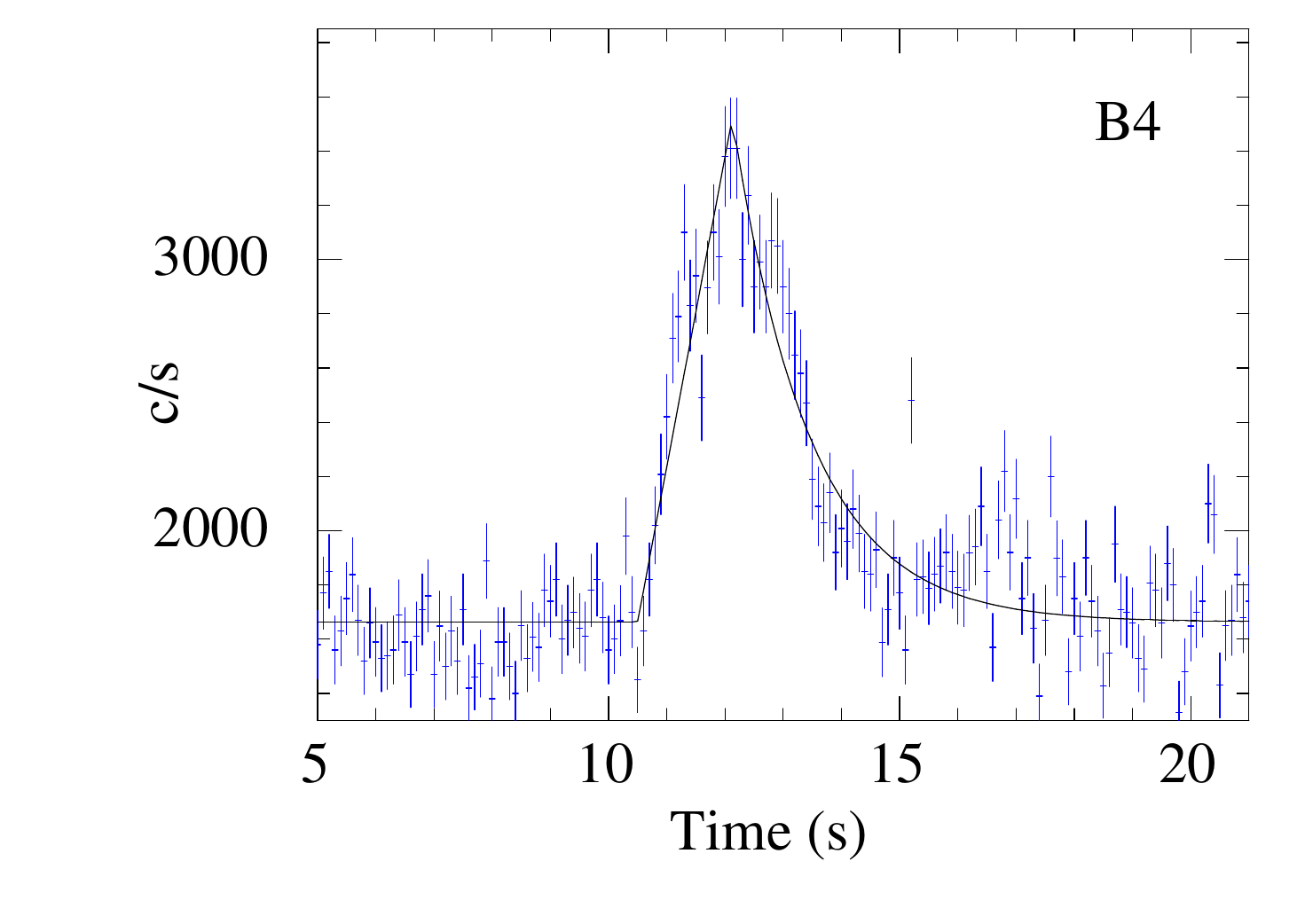}
			\includegraphics[scale=0.35,trim={5cm 5cm 8cm 6cm},clip=true,angle=0]{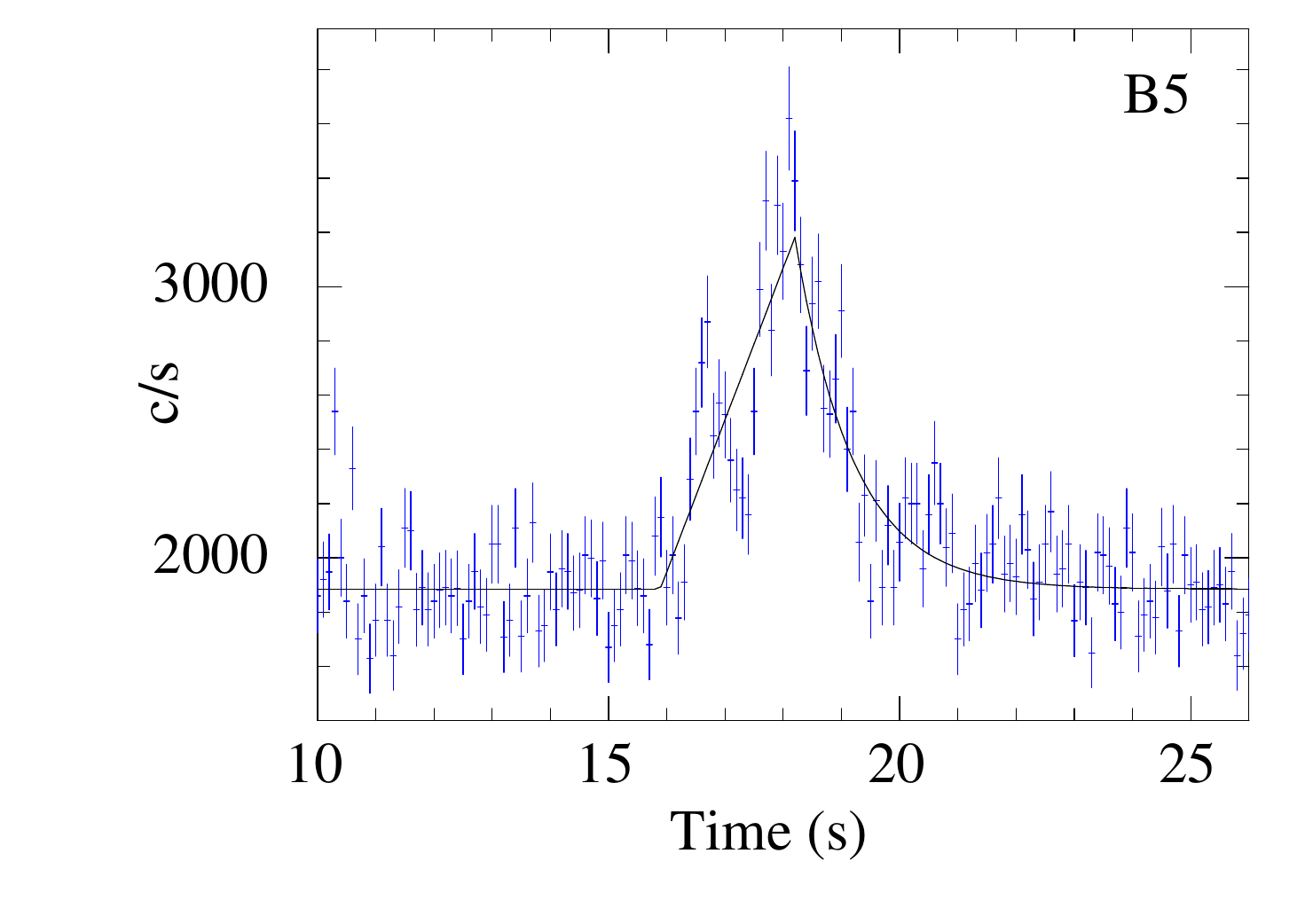}
			\includegraphics[scale=0.35,trim={5cm 5cm 8cm 6cm},clip=true,angle=0]{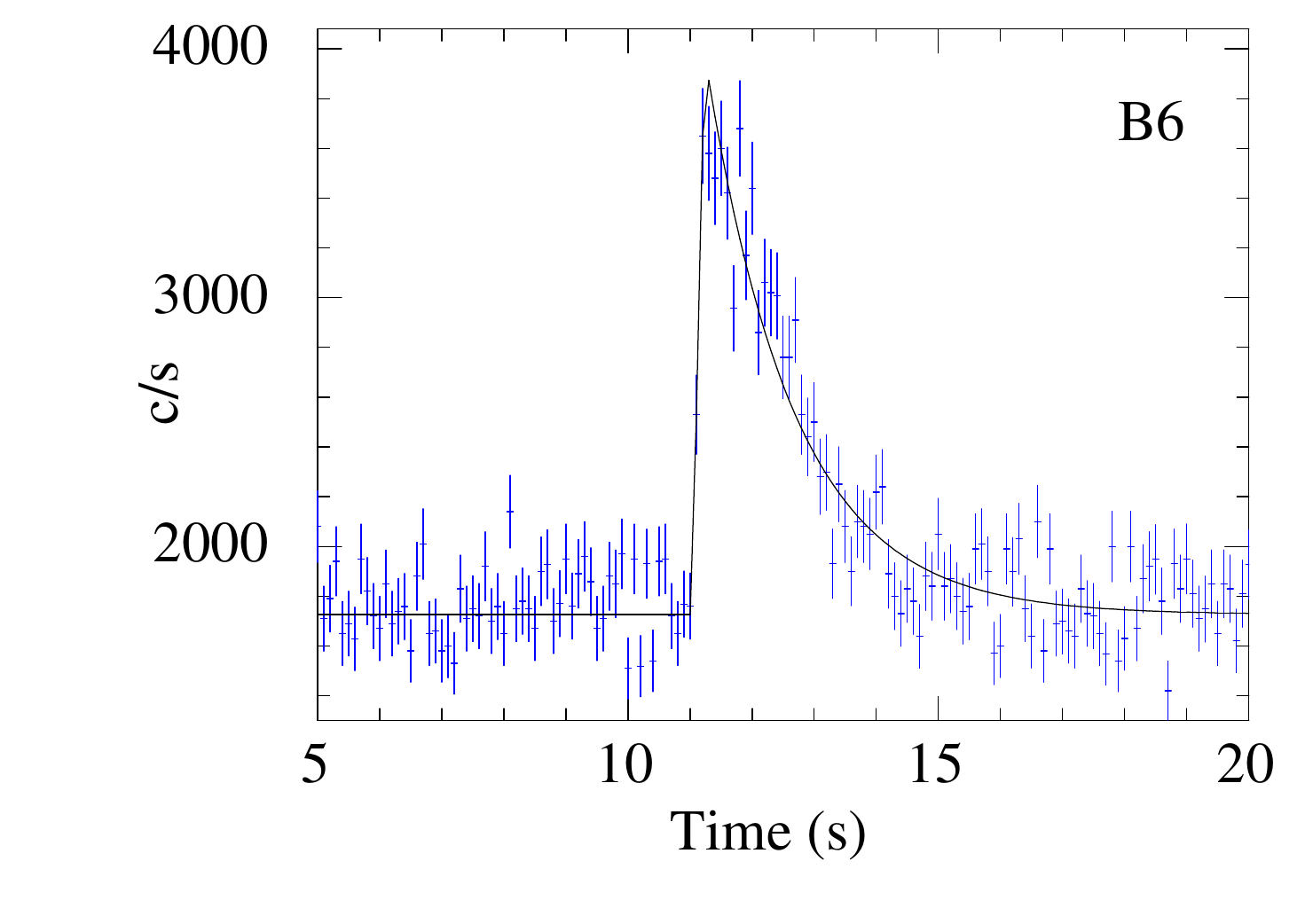}
		\end{minipage}

		\caption{Bursts detected from Cyg X-2 have been fit with a constant and a bursting model (linear rise and exponential decay). The best fit parameters are quoted in Table \ref{tab:burs}.}
		\label{fig:burst-fit}
		\end{center}
	\end{figure}

	The six bursts in the 3.0-80.0 keV energy band, were fitted using a constant plus burst model (linear rise + exponential decay) in order to obtain their timing properties, which are given in Table \ref{tab:burs}. All the bursts and burst-like  events have a FRED profile with burst durations ranging from 4-6 seconds and an average decay time of $\sim$1.2 s (see Figure \ref{fig:burst-fit}). In all the bursts, the peak count rates are seen to be around twice as that of their persistent background emission. These burst peak to persistent ratios are consistent with previously reported bursting characteristics of Cyg X-2 (see \citealt{Smale1998,Wijnands1997, Kuul1995}).
	
	\begin{table*}
		\begin{tabular}{|p{1cm}|p{2cm}|p{2cm}|p{2cm}|p{2.5cm}|p{2.7cm}|p{2.5cm}|}
			\hline
	Burst no. & Burst duration (s)& Peak to Persistent count rate ratio & Decay time (s)  & Fluence ($\times$10$^{-8}$erg/cm$^2$) & Peak flux ($\times$10$^{-9}$erg/cm$^2$/s) & BO upper limits rms frac. variability \\
			
%
			\hline
			B1 & $\sim$4 &  $\sim$2.2	& 1.24$\pm$0.14 & 1.0 &	6.3  & 0.2\% \\
			B2 & $\sim$4.5 & $\sim$2.1	& 1.38$\pm$0.13 & 1.62 & 8.9 & 1\% \\
			B3 & $\sim$6 &  $\sim$1.8	  &1.48$\pm$0.19 & 1.4 & 6.3  & 0.1\% \\
			B4 & $\sim$5 & $\sim$2.1	& 1.33$\pm$0.15	 & 1.7 & 6.7 & 1.44\%  \\
			B5 & $\sim$5 &  $\sim$1.7	 &0.99$\pm$0.19	 & 1.1 & 4.5 & 0.2\% \\
			B6 & $\sim$5 &  $\sim$2.3	 &1.42$\pm$0.13	 & 1.3 & 6.0 & 0.2\% \\
			\hline
		\end{tabular}
		\caption{Properties of all the bursts detected from Cyg X-2 using LAXPC. The burst fluence and peak fluxes have been shown for LXP10 alone in the 3.0-15.0 keV band.The upper limits to the BO rms fractional variability amplitudes have been computed in the 0.01-605 Hz frequency range in the 3.0-80 keV energy band using the light curves combined from all three detectors. }
		\label{tab:burs}
	\end{table*}

\begin{figure}[ht]
		\begin{minipage}{1.0\linewidth}
			\includegraphics[scale=0.35,trim={5cm 4cm 10cm 6cm},clip=true,angle=0]{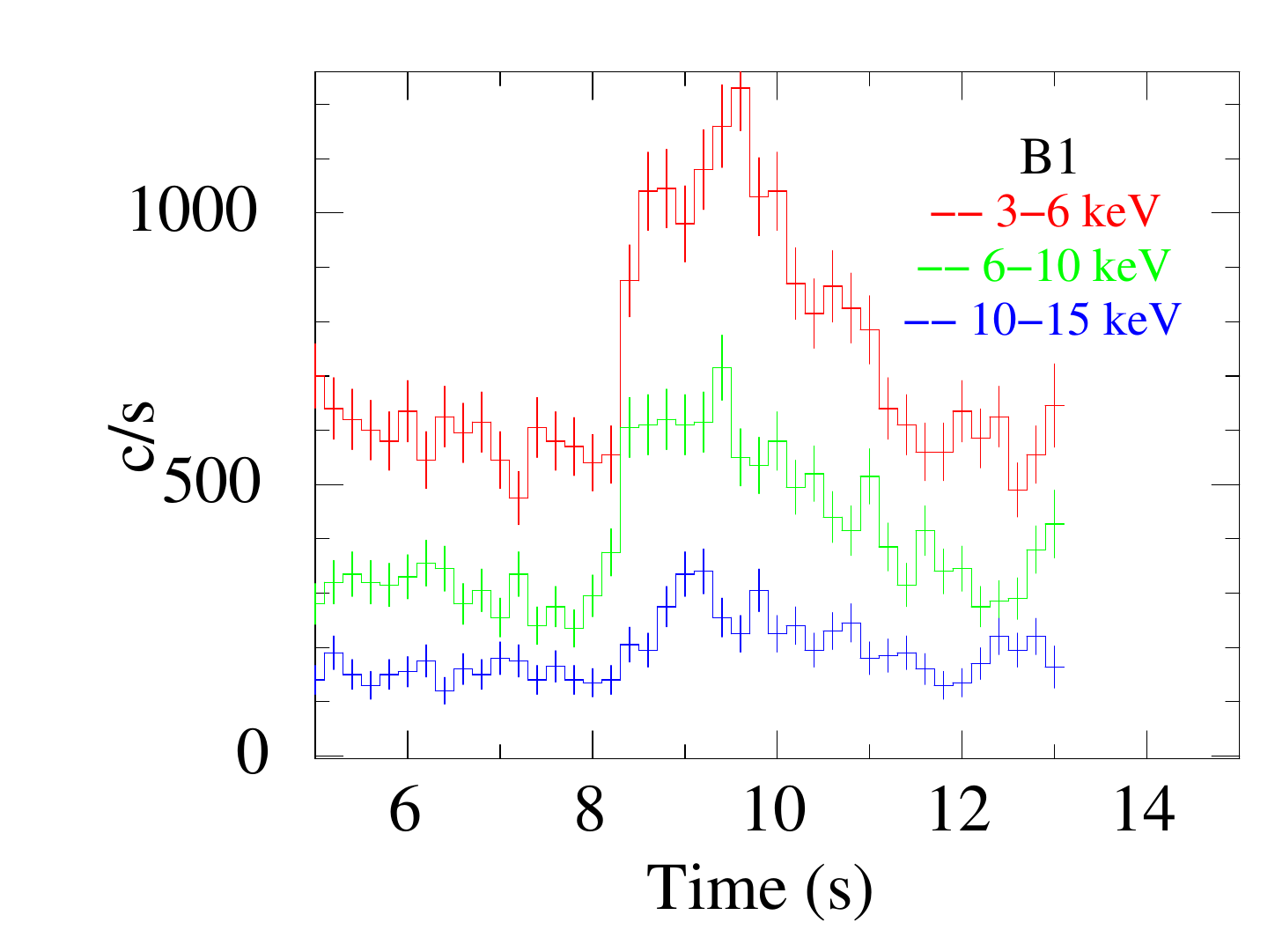}
			\includegraphics[scale=0.35,trim={5cm 4cm 10cm 6cm},clip=true,angle=0]{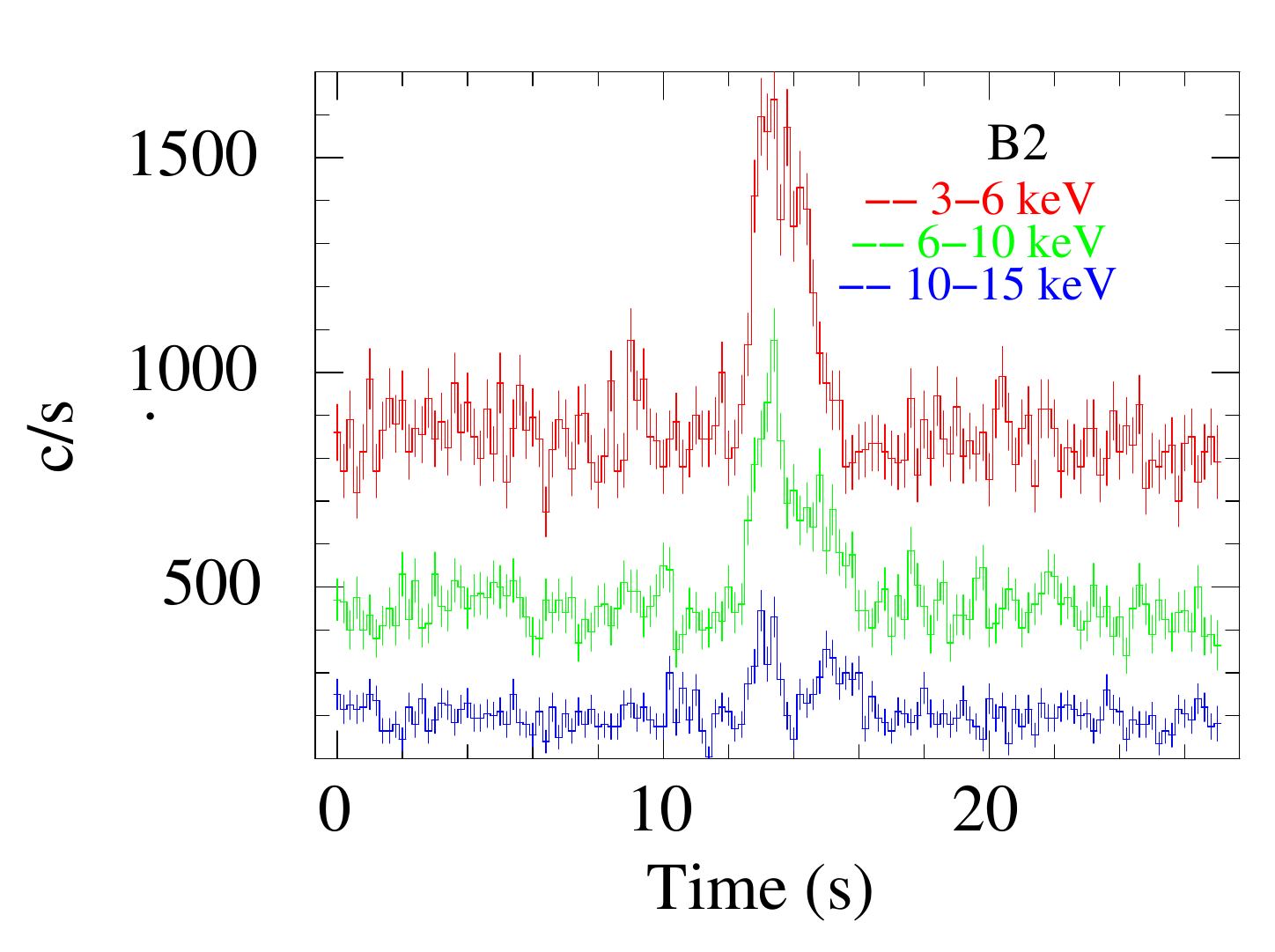}
				\includegraphics[scale=0.35,trim={5cm 4cm 10cm 6cm},clip=true,angle=0]{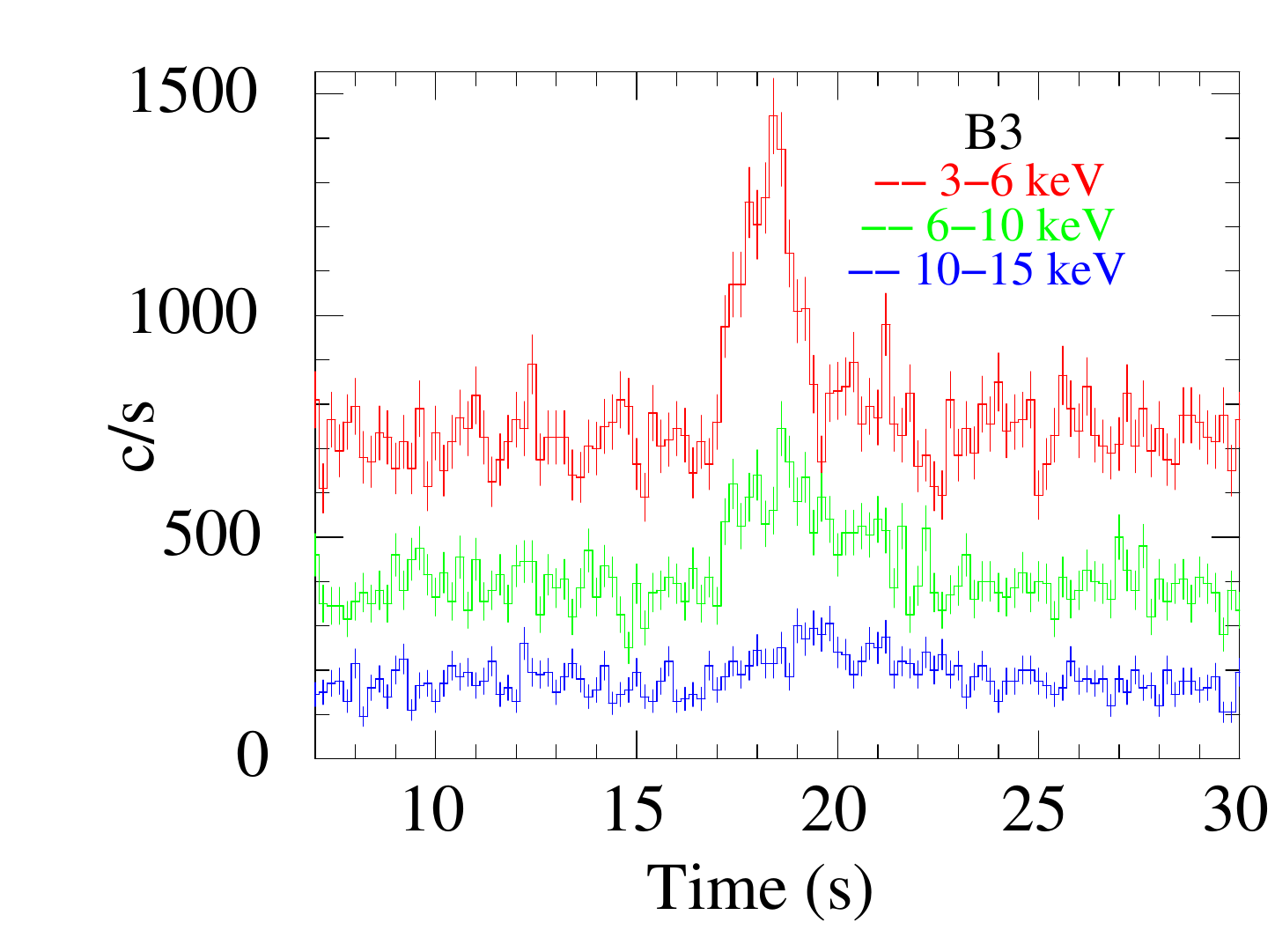}
			
		\end{minipage}     
		
		
		\begin{minipage}{1.0\linewidth}   
	\includegraphics[scale=0.35,trim={5cm 4cm 10cm 6cm},clip=true,angle=0]{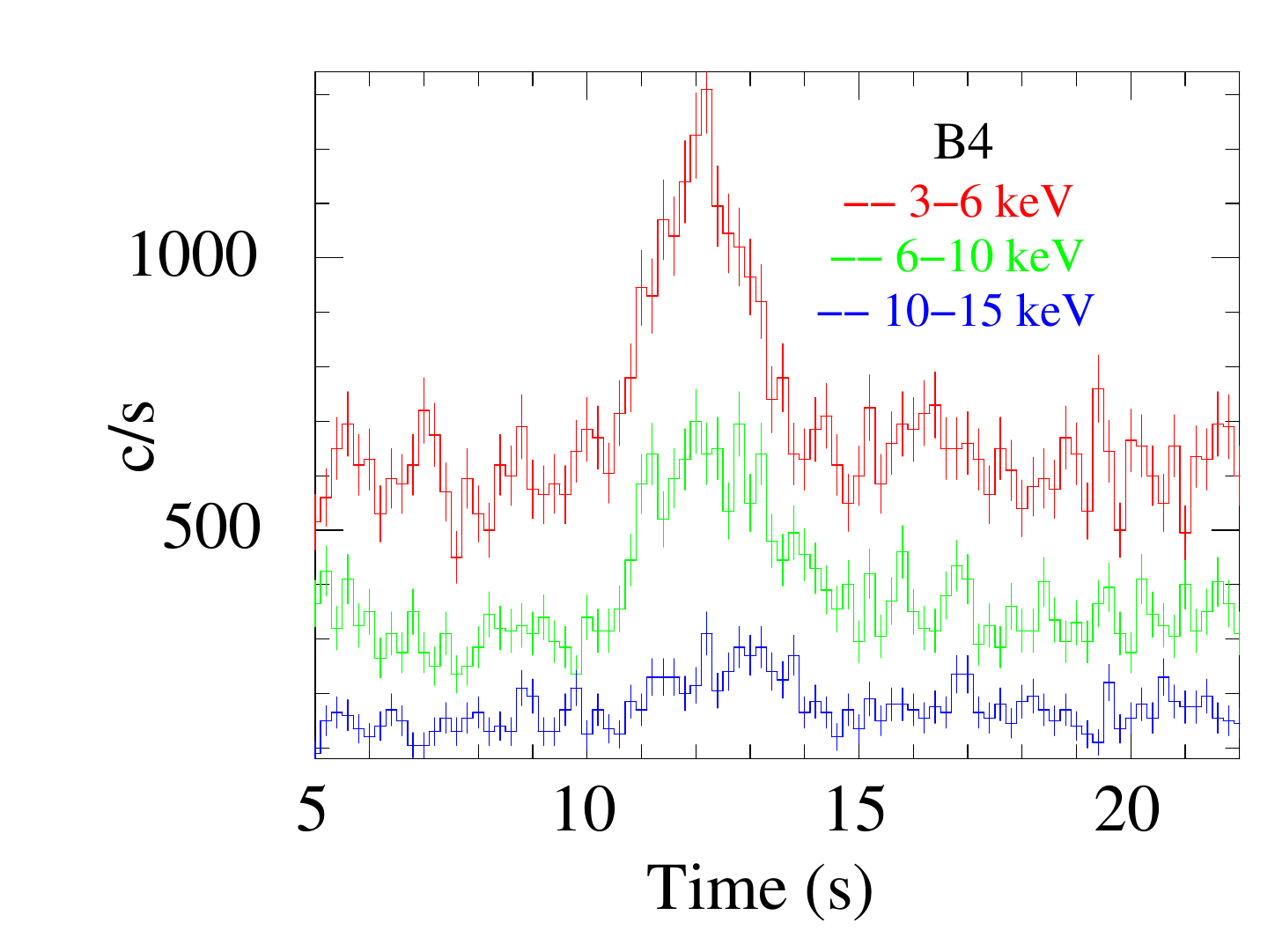}
	\includegraphics[scale=0.35,trim={5cm 4cm 10cm 6cm},clip=true,angle=0]{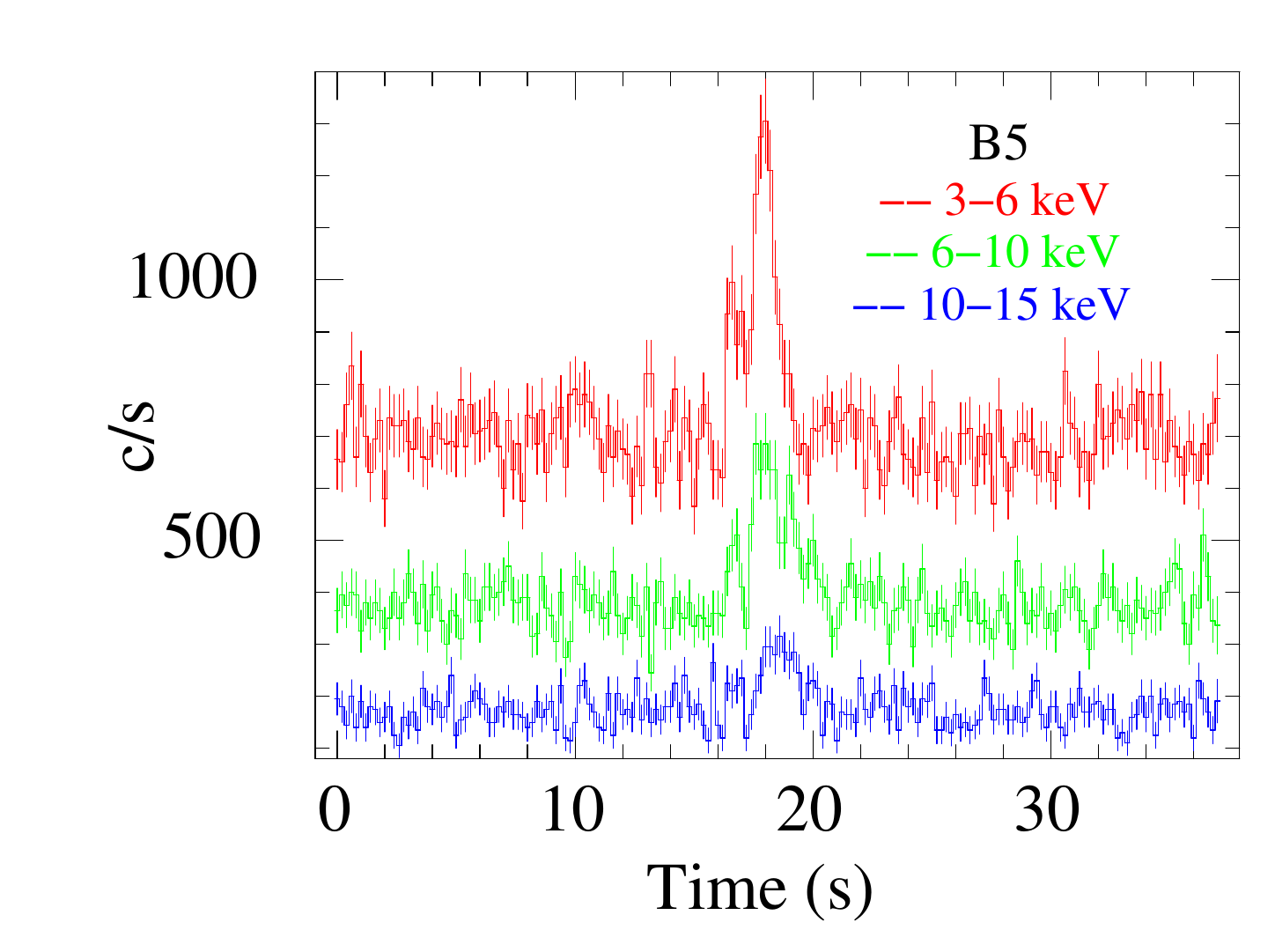}
	\includegraphics[scale=0.35,trim={5cm 4cm 10cm 6cm},clip=true,angle=0]{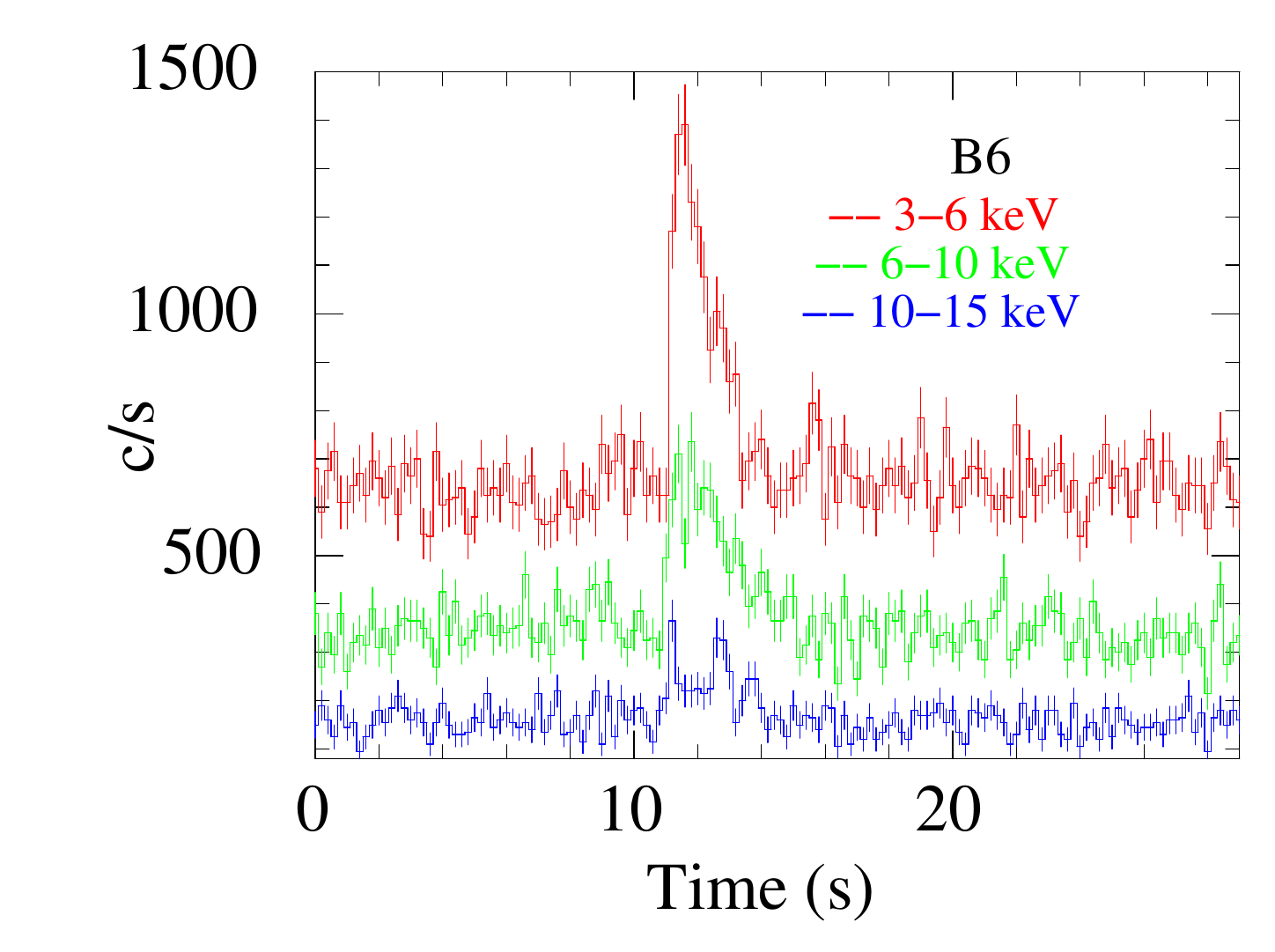}
		\end{minipage}
		
		\caption{Energy resolved burst profiles from Cyg X-2 showing a decreasing peak count rate at higher energy bands.}
		\label{fig:energyres-burs}
	\end{figure}

	Using the corresponding response files, we obtained channel to energy information for LXP10 and used that to carry out an energy resolved burst profile analysis. The burst light curves were extracted in the four energy bands: 3.0-6.0 keV, 6.0-10.0 keV and 10.0-15.0 keV. The light curves have been plotted with their true count rates in Figure \ref{fig:energyres-burs}. The burst count rates are seen to be maximum in the 3.0-6.0 keV energy band. Beyond 6 keV, as expected for Type-1 X-ray bursts, the burst peak count rates decrease.

	We also carried out a search for burst oscillations on the individual burst light curves using the GHATS timing package\footnote{http://www.brera.inaf.it/utenti/belloni/GHATS\_Package/Home.html}. We used 8192 points per FFT and a rebinning factor of 1 to generate the dynamic power spectrum with a sliding window of 1 second. The dynamic power spectra for all the bursts showed a smooth variation in power across all the frequency bins (between 1-5 kHz) as a function of time (see Appendix Figure \ref{fig:dynps}). Using the rms normalized white noise subtracted power spectra generated from the light curves combined from all the three LAXPC detectors, we derive an upper limit to the fractional rms amplitude of the BO in the 0.01 to 605 Hz range as $\sim$1\% for all the bursts (See Table \ref{tab:burs}).

	\subsubsection{Burst Spectral Analysis}
	We have also performed spectral analysis using Level 2 event files of LXP10 detector. The spectra and responses were generated using the Format A pipeline\footnote{http://astrosat-ssc.iucaa.in/?q=laxpcData}. Good Time Intervals (GTI) for each burst were specified according to their burst durations (as indicated in Table \ref{fig:burst-fit}). Source spectra were extracted corresponding to these GTI using the `\textit{laxpc$\_$make$\_$spectra}' code from the pipeline. The spectrum corresponding to the persistent emission has been considered as the background for the burst spectrum. The background spectra were thus extracted from a time segment 10 seconds prior to the onset of each burst. We have used XSPEC package \citep{Arnaud1996} for spectral fitting.

	The effective burst spectrum was fitted with a blackbody model, `blackbodyrad' in XSPEC. This model has two parameters namely, the temperature (T$_{BB}$) in keV and a normalization (K = R$_{BB}/d_{10}$), where R$_{BB}$ and d$_{10}$ are the blackbody radius and distance to the source in units of 10 kpc. We account for the line-of-sight neutral Hydrogen column density (nH) absorption by using the `Tbabs' model \citep{Wilms2000}. The burst peak fluxes and fluences are shown in Table \ref{tab:burs}. We have also added a systematic uncertainty of 1\% to the spectral fits. Since the hydrogen column density parameter is not well constrained using the LAXPC detector, we fixed the nH parameter to the average value of 0.2$\times$10$^{22}$cm$^{-2}$ as obtained from the Galactic nH calculator in the direction of the Cyg X-2 co-ordinates\footnote{See https://heasarc.gsfc.nasa.gov/cgi-bin/Tools/w3nh/w3nh.pl and references}.   
	
\begin{figure*}
		\centering
		\includegraphics[scale=0.3,angle=0,trim={5cm 1cm 5cm 2cm}]{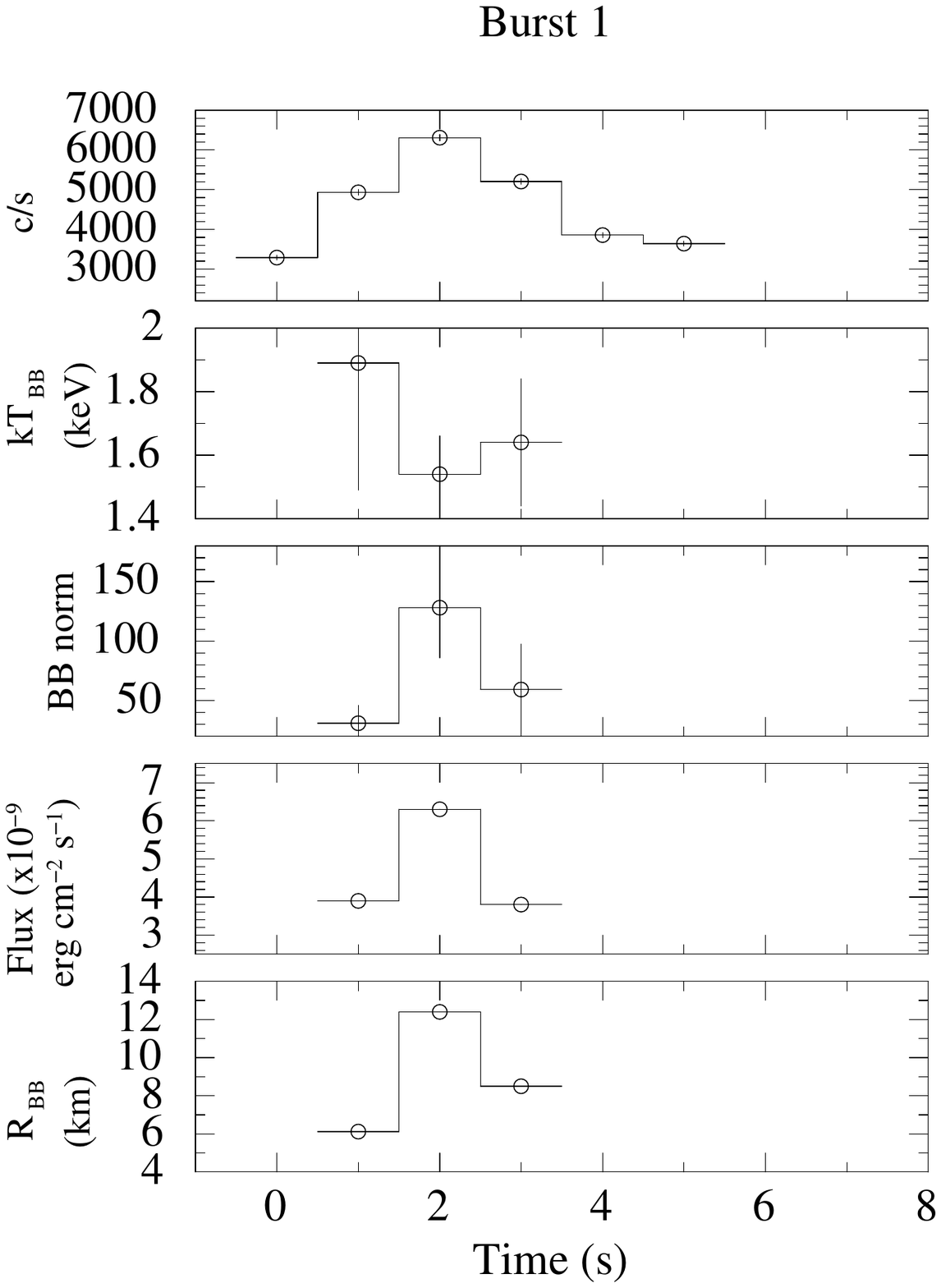}
		\includegraphics[scale=0.3,angle=0,trim={8cm 1cm 5cm 2cm}]{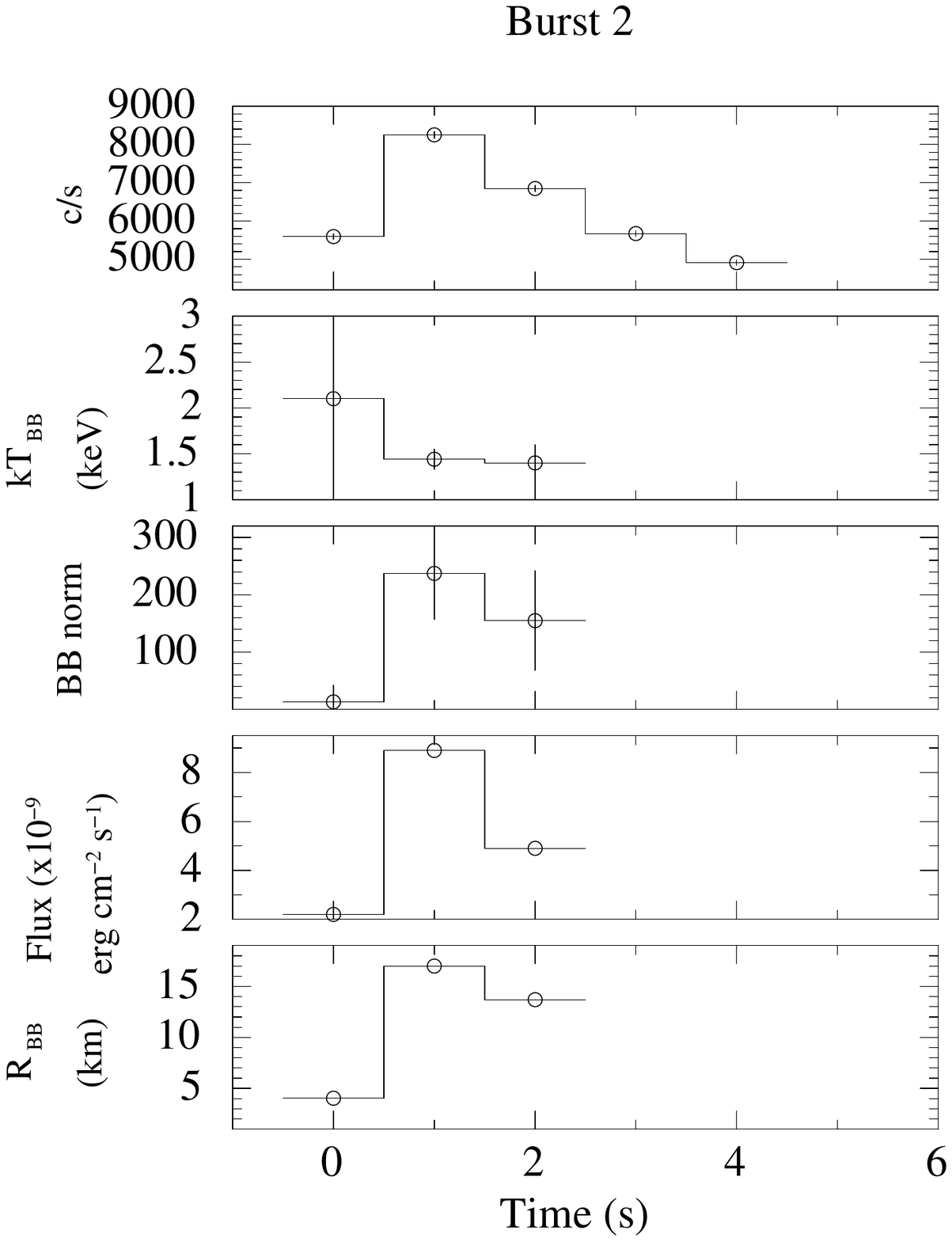}
		\includegraphics[scale=0.3,angle=0,trim={8cm 1cm 5cm 2cm}]{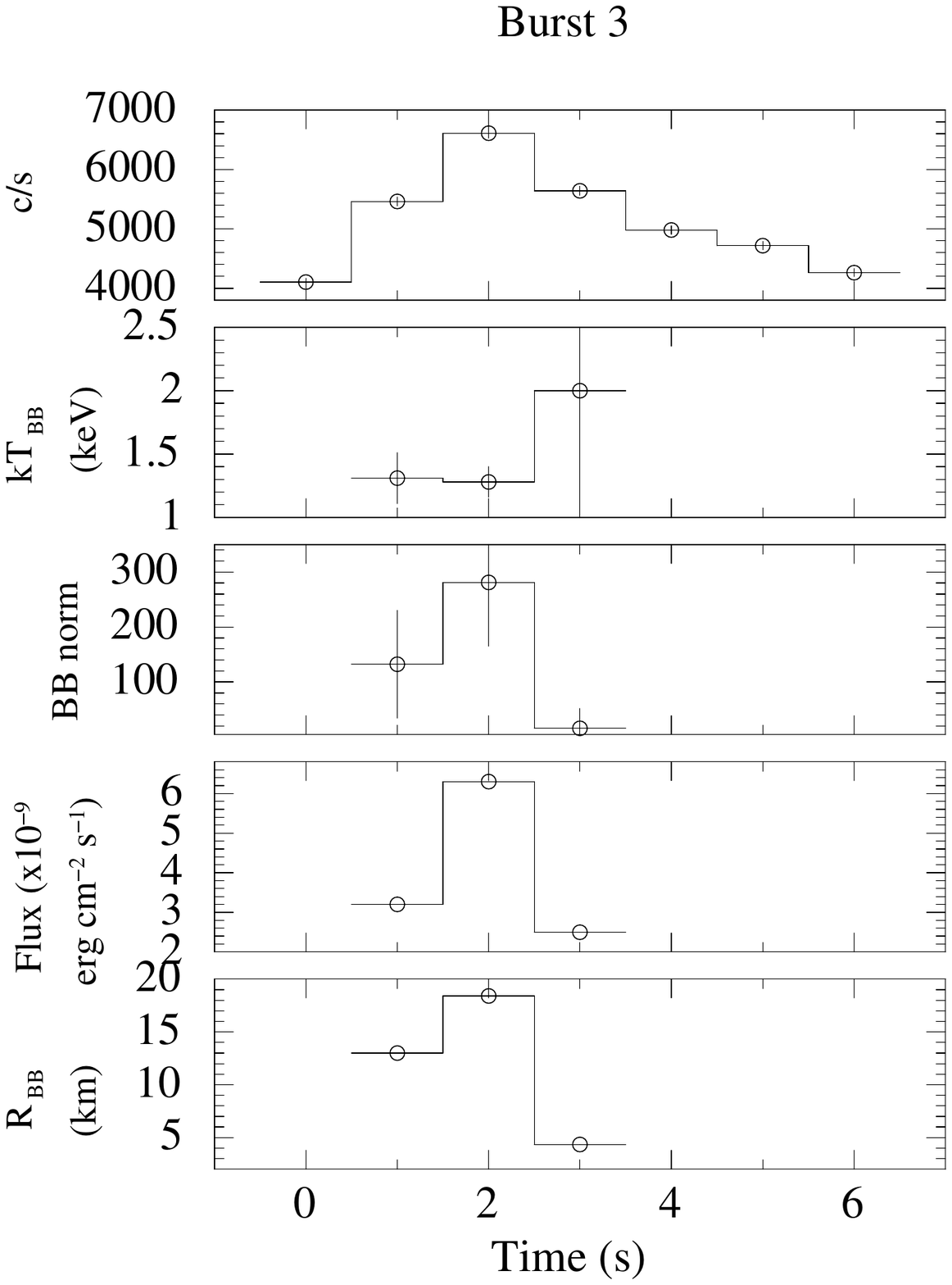}
		
		\includegraphics[scale=0.3,angle=0,trim={5cm 1cm 5cm 1cm}]{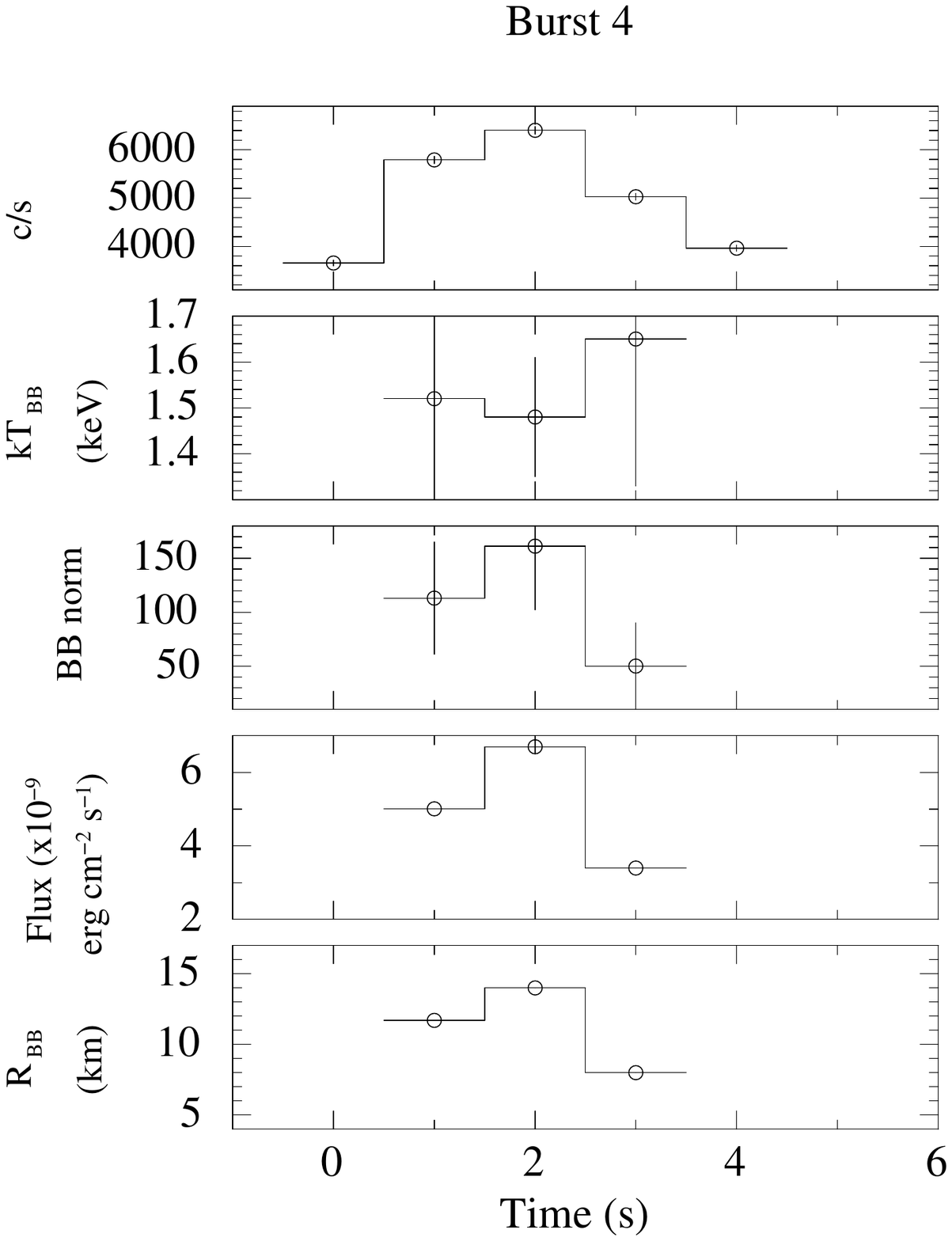}
		\includegraphics[scale=0.3,angle=0,trim={8cm 1cm 5cm 1cm}]{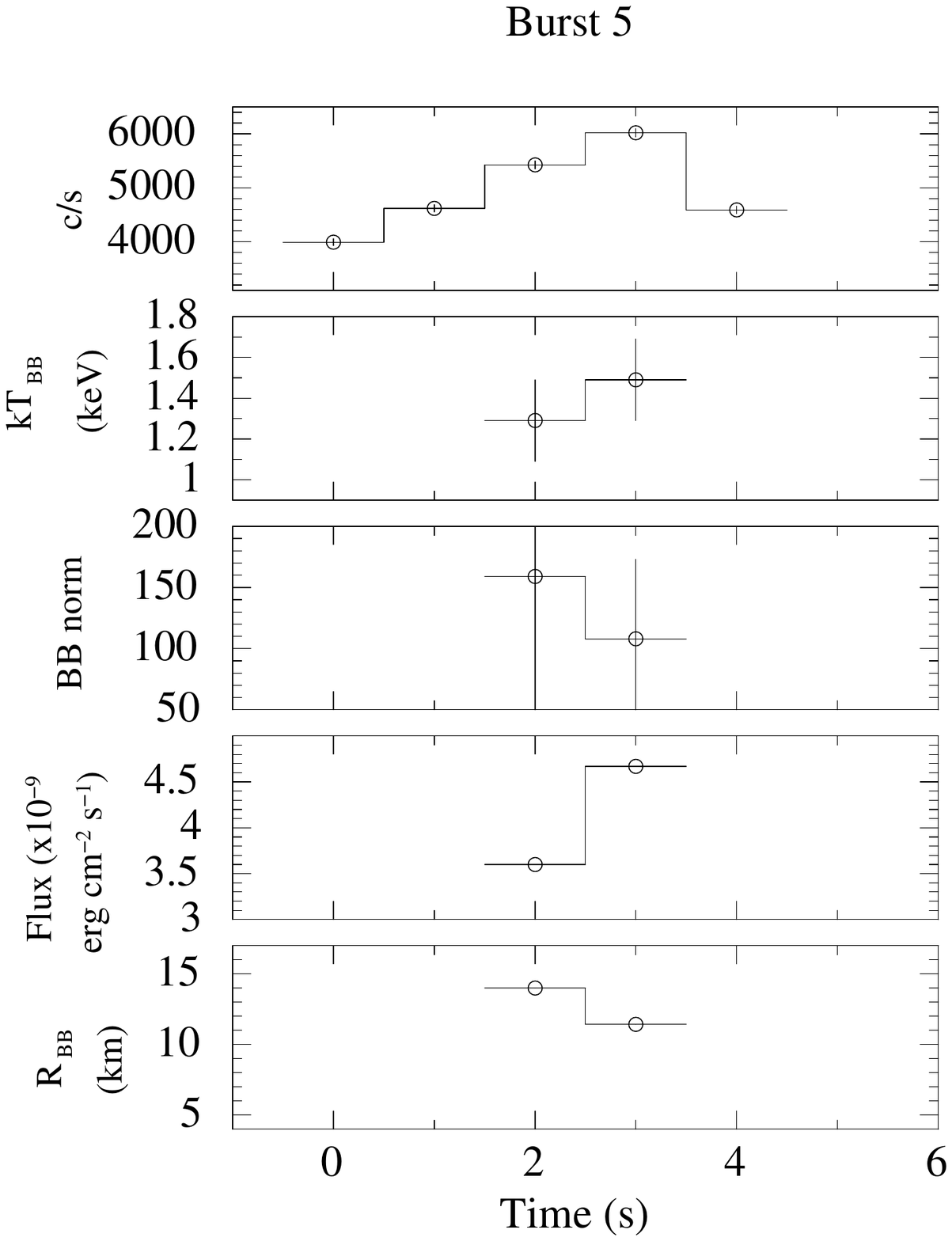}
		\includegraphics[scale=0.3,angle=0,trim={8cm 1cm 5cm 1cm}]{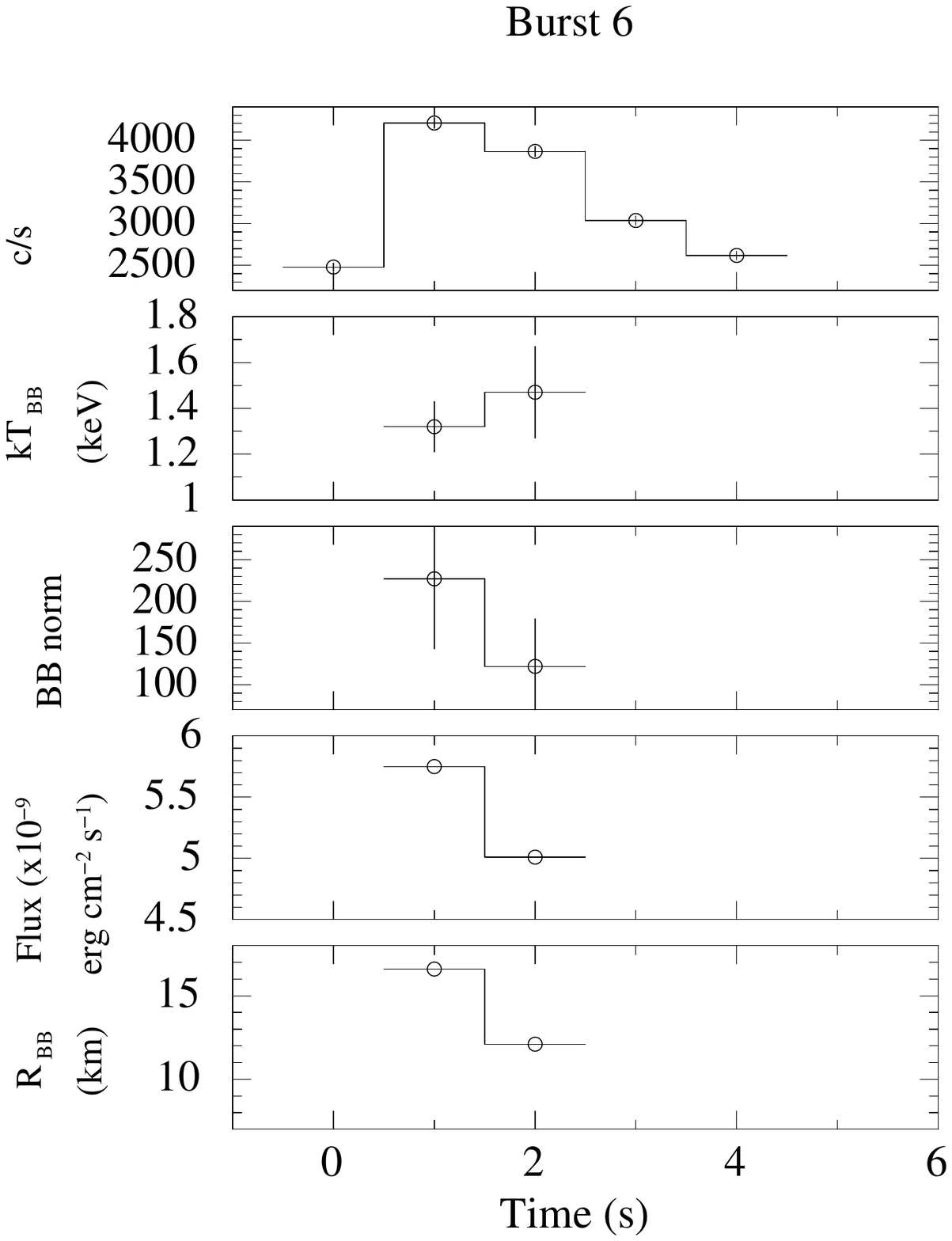}

		\caption{Variation of the best fit spectral parameters in the time resolved burst spectral analysis for the six bursts detected from Cyg X-2 is shown as a function of time.}
		\label{fig:burst-time-res}
	\end{figure*}

	We further carried out a time resolved burst spectral analysis for each of the bursts with a time slice of 1 s. Each of the burst segments were modeled using the same time averaged burst spectral model, i.e. \textit{`TBabs*bbodyrad'}. The persistent spectrum corresponding to the 10 second long time segment prior to the onset of each burst is again taken as the background spectrum. Some of the segments did not have enough statistics in order to carry out a meaningful fit, these have been ignored. The burst spectra were statistically significant only in the spectral range 3.0-20.0 keV as also observed in the time resolved burst light curves (see Figure \ref{fig:burst-time-res}). 
	The evolution of the blackbody temperature and radius along with the flux as a function of time  is shown in Figure \ref{fig:burst-time-res} for all the six bursts. All of the bursts exhibit a blackbody temperature in the range 1-2 keV while the radius is seen to vary widely between 5-18 km. In some bursts (like B1, B2, B3 and B4), the peak temperature is observed to be lower than the adjacent bins. Similarly, the blackbody radius and flux is observed to increase towards the peak and drop down post the peak. These trends in the variations are however poorly constrained due to large errors and lack of parameter coverage throughout the burst. We therefore refrain from conclusively commenting on whether these trends indeed indicate a PRE behavior.

	\subsection{Non-burst analysis}
	
	\subsubsection{Colour-Colour diagrams and Hardness-Intensity analysis}
	
\begin{figure}
		\centering
		\includegraphics[scale=0.38,angle=-90,trim={2.2cm 2cm 2cm 2cm},clip=true]{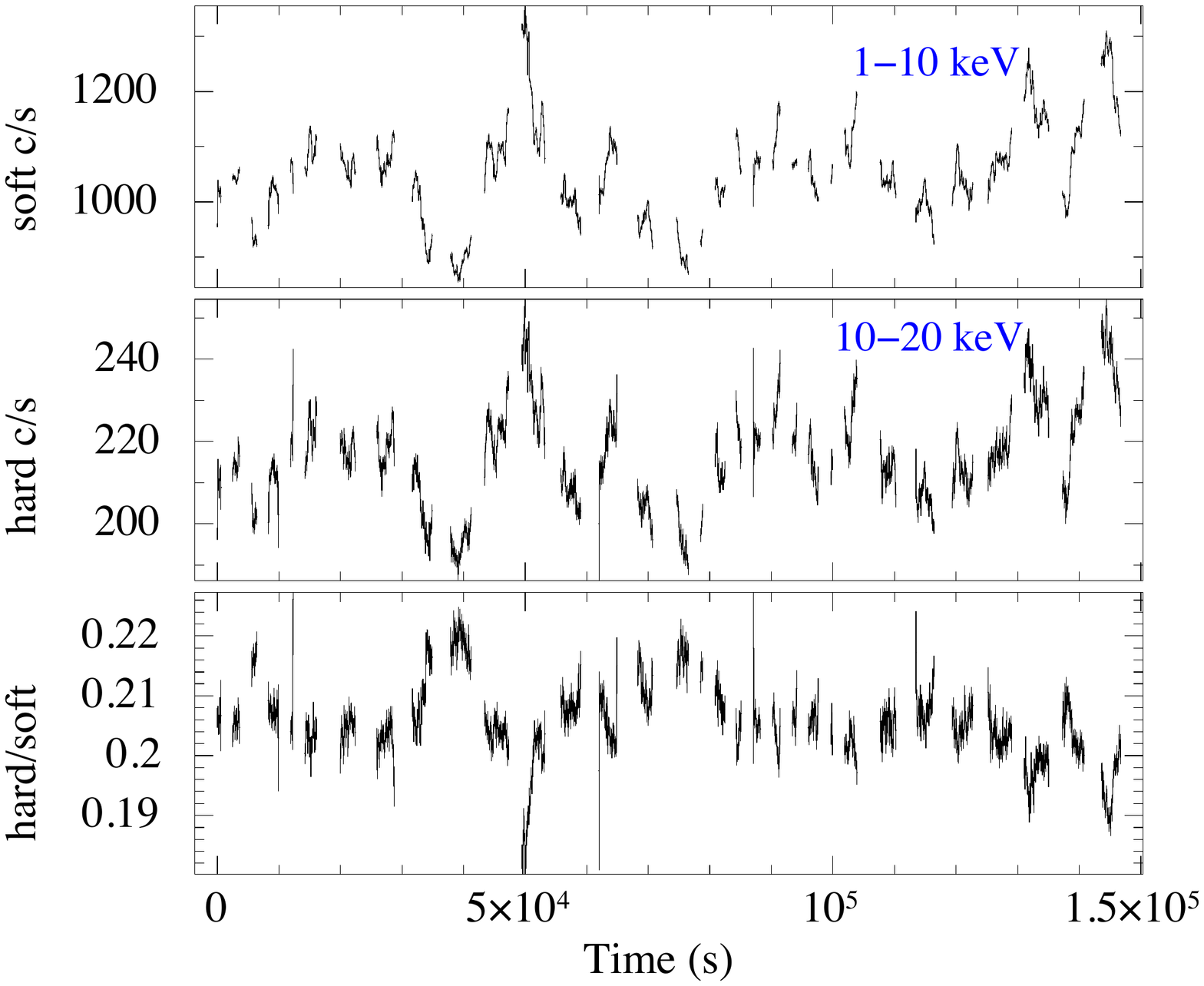}
		\caption{Energy resolved light curves from Cyg X-2 in the soft (3.0-10.0 keV) and  hard (10.0-20.0 keV) energy bands. The bottom panel shows the hardness ratio. }
		\label{fig:h-ratio}
	\end{figure}

	The bursting segments were windowed and removed in order to study the persistent emission and characterize the source state. A sample of the windowed light curve from LXP10 is shown in Figure \ref{fig:h-ratio}. The X-ray light curve in both energy bands shows significant variability around a persistent mean along with some dips. 
	
	To characterize the spectral variation and hardness ratio as a function of intensity, we extract the light curve in different energy bands. The LXP10 light curves in 3.0-10.0 keV and 10.0-20.0 keV are shown in the top two panels of Figure \ref{fig:h-ratio}. The soft X-ray photon count rates are higher by a factor of 6 as compared to the hard X-ray count rate. The bottom panel of Figure \ref{fig:h-ratio} shows an increase in the hardness ratio during the dipping epochs. Such a behavior is common with several large inclination dipping LMXB sources (for example EXO 0748-676, XTE J1710-281, etc.).

\begin{figure*}
		\centering
		\includegraphics[scale=0.2,angle=0,trim={0cm 1cm 0.0cm 0.0cm},clip=true]{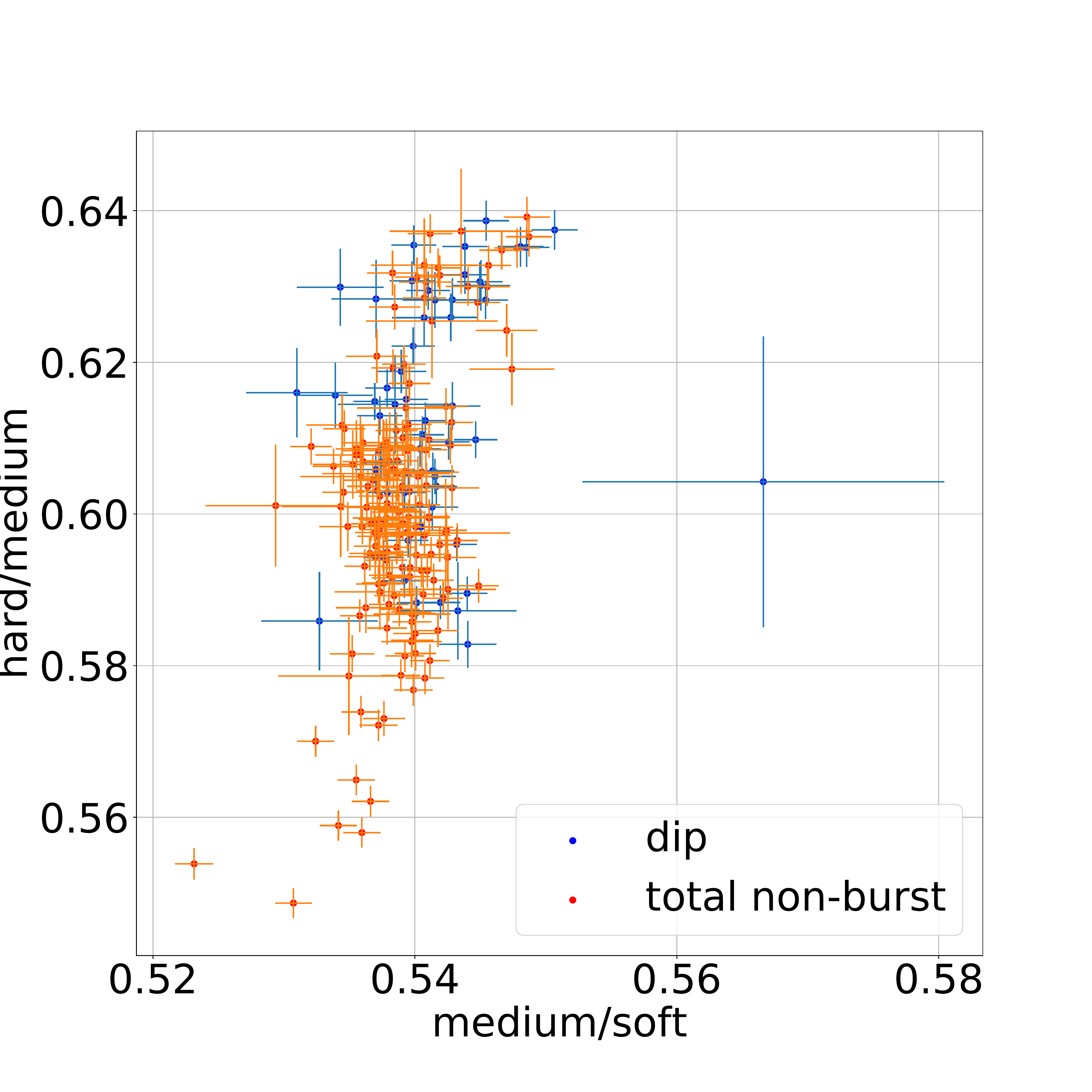}
		\includegraphics[scale=0.2,angle=0,trim={0cm 1cm 3.5cm 0.0cm},clip=true]{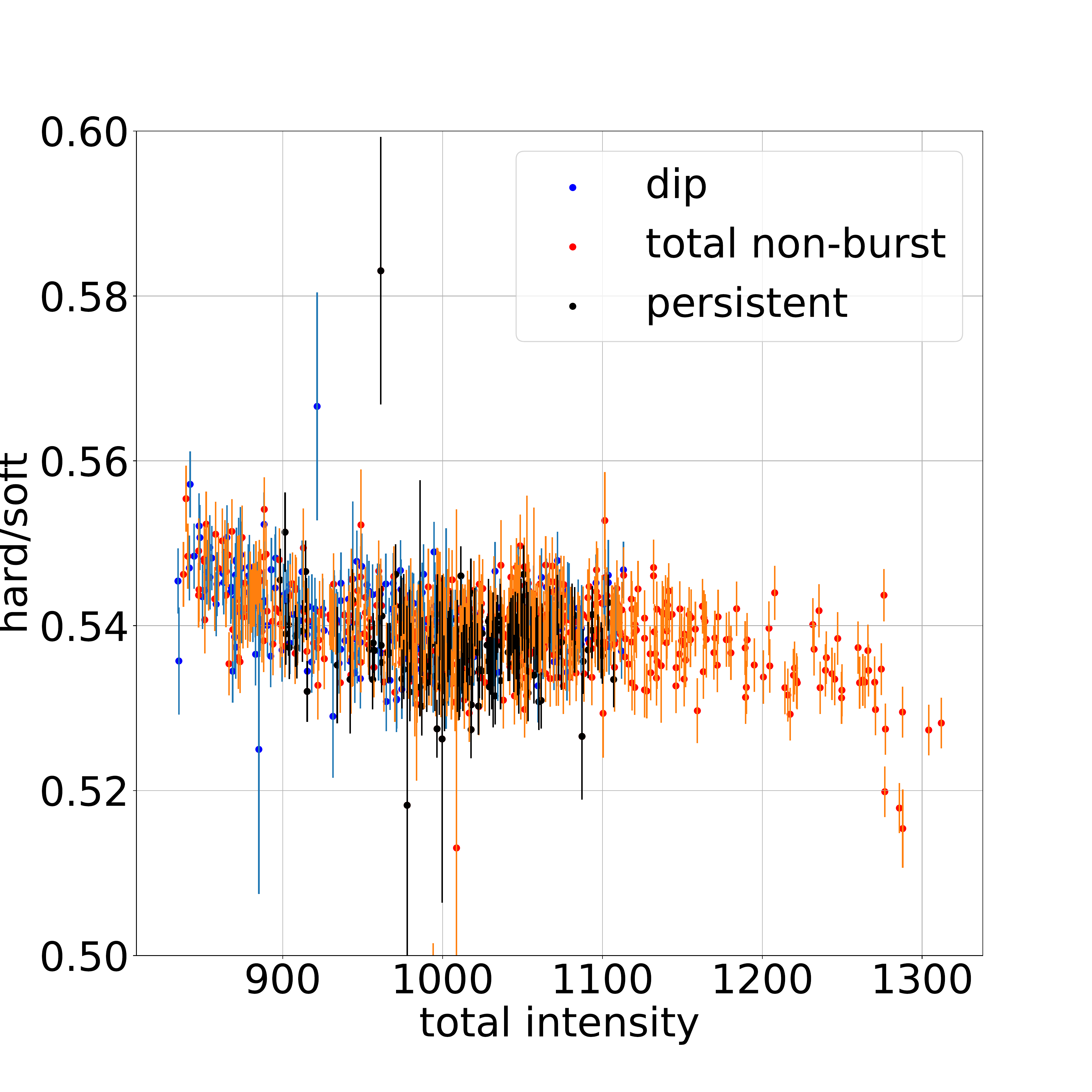}
		\caption{The colour colour diagram (left) and the Hardness intensity diagram (right) constructed for Cyg X-2 using the dipping, persistent segments as well as the total non-burst LXP10 light curves in the 3-6 keV, 6-10 keV and 10-15 keV energy bands.}
		\label{fig:ccd-hid}
	\end{figure*}

	We further construct the Colour Colour Diagram (CCD) and Hardness-Intensity Diagrams (HID) using the non-burst LXP10 light curves in  energy bands: 3-6 keV, 6-10 keV and 10-15 keV (See Figure \ref{fig:ccd-hid}). The ratios that are defined as soft and hard colour, are shown in the Figure axes. The HID and CCD have been constructed individually for the dipping segments, persistent emission and the total dip plus persistent segments and have been overlayed together. In the left panel of Figure \ref{fig:ccd-hid}, the dipping and overall non-burst (persistent+dip) segments both indicate moderately hard photon ratios. In the right panel, the hardness ratio as a function of the total intensity clumps around $\sim$1000 and around $\sim$900 for the persistent and for the dipping segments, respectively. The most striking feature is that the hardness ratio remains steadily at around 0.54 across all non-burst source intensities. The hardness-intensity and colour ratios have been compared with previous reports of the different states. The FB and HB are defined such that they have a constant hardness ratio across varying intensities, with the FB being less hard compared to the HB. The NB exhibits a varying hardness ratio with varying intensity. The horizontal structure observed in our HID can indicate either the HB or the FB depending upon the hardness ratio levels. To further resolve the nature of the branch, we carry out a persistent emission spectral analysis followed by analysis of the power density spectrum (PDS).

	\subsubsection{Non-bursting spectral analysis}
	
\begin{figure}
		\centering
		\includegraphics[scale=0.35,angle=-90,trim={0cm 0cm 0cm 0cm},clip=true]{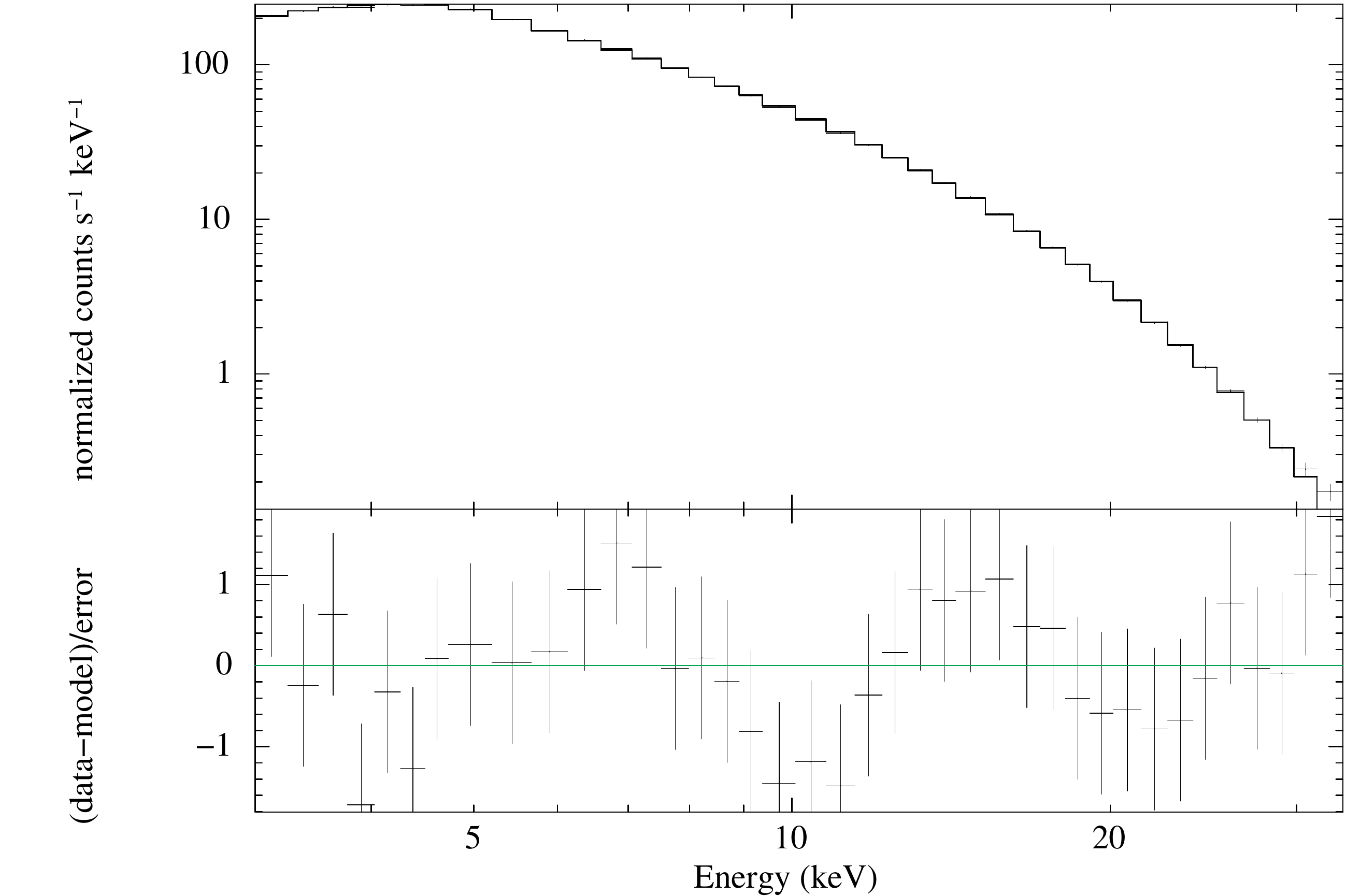}
		\caption{Non-bursting spectral fit for LXP10 persistent emission from Cyg X-2 in the 3.0-32.0 keV energy band.}
		\label{fig:spec}
	\end{figure}

	We carried out a non-bursting spectral analysis to identify the continuum emission and possibly extract information about the spectral branch of the source. We created source and background spectra using GTI to exclude the bursting epochs. The extraction was done using the `Format A' AstroSat-LAXPC pipeline using standard filtering of the SAA and Earth occultation epochs. The LXP10 persistent spectrum was fit in XSPEC with an added systematics of 1\%. The spectrum beyond 32 keV was dominated by background and so we have ignored those energies. We fit the 3.0-32.0 keV continuum spectrum using a thermal comptonization model `compTT' \citep{Titarchuk1994} which represents Comptonization of soft photons in a hot plasma. This is used along with the  Tuebingen-Boulder ISM line-of-sight neutral hydrogen absorption model `TBabs' which utilizes the updated photoionization cross sections as well as abundances \citep{Wilms2000}. We fixed the nH parameter to the galactic value of 0.2$\times$10$^{22}$/cm$^2$. We obtained an excess in residuals below 10 keV and a high reduced $\chi^2$ of 2.6 ($\chi^2$ of 87 for 39 d.o.f). This prompted us to include the disk blackbody model along with the compton model. We obtain  best fit parameters as follows: disk blackbody temperature of T$_{in}$=2.0$\pm$0.14 keV, input soft photon Wein temperature (T$_0$)=0.31$\pm$0.2 keV and the plasma temperature (kT)= 3.4$\pm$0.12 keV. We obtain a reduced $\chi^2$ of 0.92 with 39 d.o.f (Figure \ref{fig:spec}). The overall 3.0-32.0 keV band flux in the non-burst persistent emission is found to be $\sim$4.9$\times$10$^{-9}$ergs/s/cm$^2$ and $\sim$2.7$\times$10$^{-9}$ergs/s/cm$^2$ for the comptonized component and the blackbody component, respectively. Assuming a distance of 11 kpc we obtain a L$_{compt}$ of 7.2$\times$10$^{37}$ergs/s and L$_{bbody}$ of 2.1 $\times$10$^{37}$ergs/s; for a distance of 8 kpc, we obtain a L$_{compt}$ of 3.8$\times$10$^{37}$ergs/s and L$_{bbody}$ of 2.1 $\times$10$^{37}$ergs/s.  These luminosities indicate that the source has reached 0.4$\times$L$_{Edd}$ (assuming a 1.4~M$\odot$ neutron star) during these observations in 2016. This could indicate far NB to an early onset of the FB. 

	\subsubsection{Power density spectrum}
	
	\begin{figure}
		\centering
		\includegraphics[scale=0.4,angle=-90,trim={0cm 1cm 0cm 0cm},clip=true]{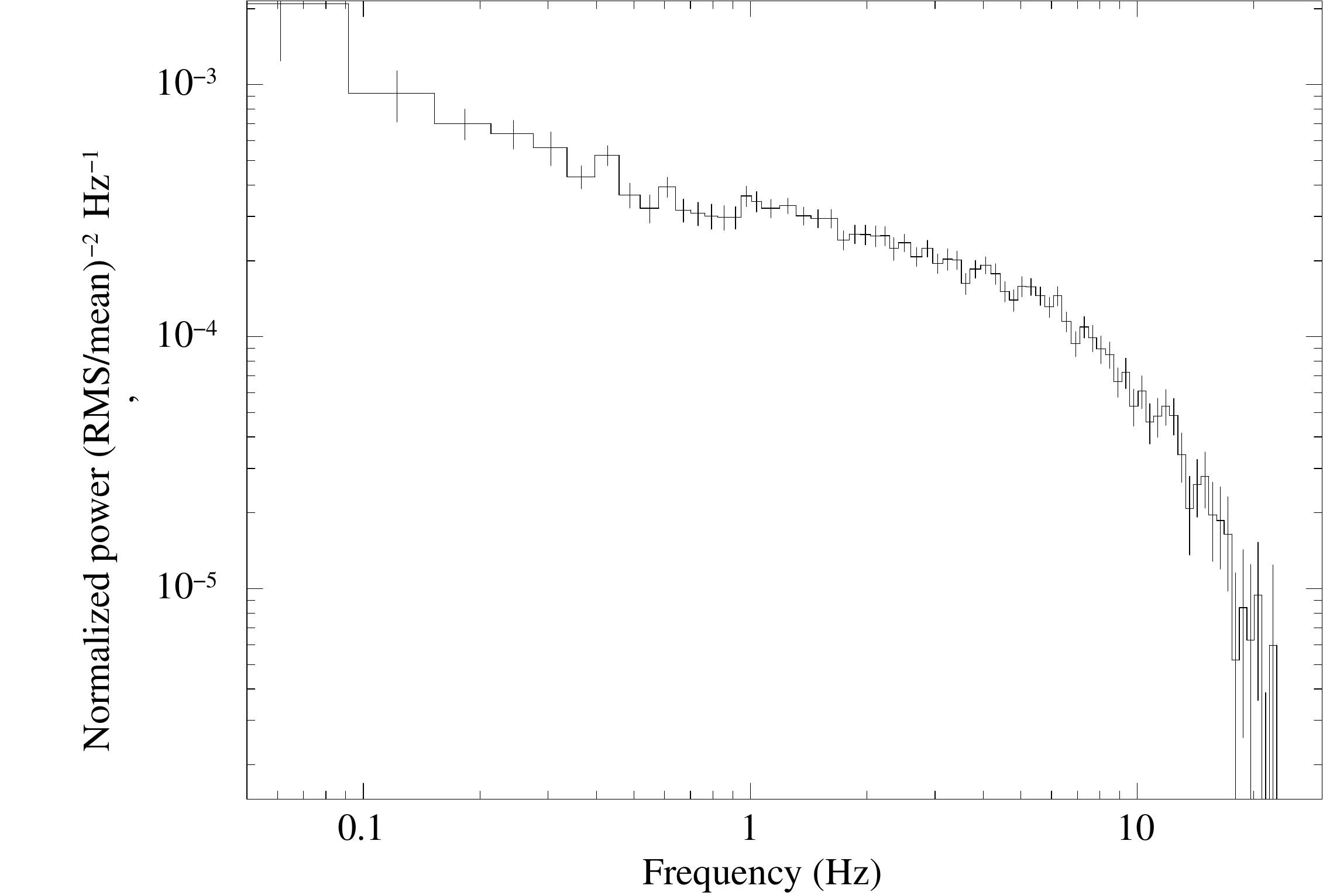}
		\caption{White noise subtracted and RMS normalized power density spectrum obtained for the LXP10 time series of Cyg X-2 binned at 0.1 ms, excluding the bursting segments. There is a steep drop in the power spectrum after 20 Hz.}
		\label{fig:pds}
	\end{figure}
	
	One crucial indication (and our final line of examination) of the source spectral branch is the variability signatures in the power density spectra. To this effect, we carry out a power spectral analysis on the light curve that excluded the bursts, using the ftool \textit{`powspec'} in HEASoft. Our primary aim was to look for either slow ($\sim$Hz) or high frequency (300-900 Hz)  QPOs in the PDS that would distinguish the HB from the FB. We obtained a featureless smooth power spectrum (see Figure \ref{fig:pds}). No significant periodicity was observed at any particular frequency.  We further carried out a search for possible QPOs in the power density spectrum around 5.6 Hz (previously reported by \citealt{Wijnands1997}). Since the power drops significantly post 20 Hz, we were unable to carry out searches in the high frequency regime. We do not detect any QPOs in the 3.0-80 keV PDS and obtain an upper limit to the fractional rms variability around 5.6 Hz as 3.4\%. The presence of high frequency QPOs is a direct indication of the HB \citep{Hasinger1987} and the lack thereof can point towards a FB.\footnote{We eliminate the possibility of the source being in the NB mainly due to lack of QPOs as well the HID-CCD shape.}   
	
	\section{Discussion and conclusions}
	
	Cyg X-2 is a highly interesting LMXB system. Even within the class identified as Cyg-like sources, Cyg X-2 particularly stands out for showing unique timing and spectral features along the Z track. Since the time of its discovery, it has been showing several intensity states, secular intensity variations, thermonuclear burst-like events and QPOs. We have carried out timing and spectral analysis of this object during onset of the FB  (left NB/FB soft vertex in the Z shape) and have identified 6 bursts/burst-like events using the LAXPC 2016 observations. A search for BO has resulted in a null detection up to 5 kHz in all the 6 bursts. We also observe the dipping segments with correspondingly higher hardness ratios in the X-ray light curve. Our timing and spectral analysis of the bursts and non-burst emission indicate that the object was in the early Flaring branch (FB) during the 2016 observations.

	Type-1 thermonuclear X-ray flashes are a common phenomenon that occur on the surface of neutron stars, during which X-ray intensities shoot up by almost an order of magnitude as compared to the persistent intensity levels. They occur as a result of unstable thermonuclear burning of fuel comprising of Hydrogen and/or Helium (see \citet{LewinVanTaam1995} and references for an overview). The different regimes of stable and unstable burning are a function of the mass accretion rate (see \citealt{Bildsten2000} for more details). Using the 3.0-32.0 keV flux of 4.9$\times$10$^{-9}$ erg/s/cm$^2$, we inferred a source luminosity of 7.2$\times$10$^{37}$ erg/s. Assuming that accretion is significantly responsible for the observed luminosity, we calculate a  mass accretion rate of $\dot{M}_{acc}$ = 4$\times$10$^{17}$g/s  ($\sim$5.9$\times$10$^{-9}$M$\odot$/yr). At accretion rates exceeding a critical value of  2$\times$10$^{18}$g/s, Helium bursts are known to be completely suppressed \citep{KG1984}. However, if the accretion rates are close to such a critical value, as should be the case during the early onset of the FB, small perturbations can lead to occasional flashes that can be weak (see for example, \citealt{KG1984}), which  would give rise to Helium dominated bursts like the ones observed in the present work. Helium burning occurs rapidly via strong interactions, unlike the slower beta decay process that occurs during Hydrogen burning \citep{vanP1998,Cumming2004}. They are characterized by short rise times ($\sim$1s) and short overall durations, often accompanied by PRE as well \citep{LewinVaccaBasinka1984}. The rapid bursts/burst-like events observed in this work are of short duration ($\sim$5 second long) and recur at timescales of a few hours. PRE burst spectra also typically yield an expanded blackbody radius and a reduced effective temperature with an approximately constant bolometric luminosity \citep{Zac_thesis_2020}. Earlier timing studies of Cyg X-2 have even shown evidence of the short duration Helium fueled bursts exhibiting photospheric radius expansion \citep{Smale1998,Tit2002}. Results from time resolved burst spectral analysis in this work indicates, although with low statistical certainty, that the blackbody radius and temperature undergo an evolution along the burst in a manner similar to a PRE. We caution that some of our parameter estimates have large uncertainties due to contamination from the persistent emission. In a previous study of Cyg X-2 using RXTE data, typical blackbody radii  were reported to be of the order of 10 - 12 km  during the FB, and $\sim$ 5 km during the HB (see \citealt{Church2010}). This is consistent with the blackbody radii observed in this paper. Results from the present work also indicate that the burst peak count rate is about 2-2.5 times the persistent count rate, consistent with previous burst reports in this source (see \citealt{Smale1998,Kuulkers1999}). Such peak-to-persistent intensity ratios are indicative of medium-to-high levels of accretion that could itself be responsible for suppressing the burst peak intensities \citep{KG1984}.

We also observe an energy dependent behavior in the Type-1 X-ray burst peak count rates, where the bursts are prominent at softer energies (between 3-6 keV) and decrease in strength and intensity at higher energies. Burst photons are predominantly soft X-ray thermal photons. Particularly, the burst-like event B5 shows a double peak where the first short peak looks like a failed burst, followed by a more fully-loaded burst. Such flashes are a common occurrence during intermediate-high level accretion rates in the early FB \citep{Zhang2009}.

The Z-track sources are the brightest objects from the LMXB class of sources that accrete persistently near their Eddington limit. Unlike the ``Sco-like" Z sources (for eg. Sco X-1, GX 17+2 and GX 349+2) which exhibit short and vertical HB and an extended FB, the ``Cyg-like" sources (like Cyg X-2, GX 340+0, GX 5-1) exhibit a horizontal HB and a weak FB  \citep{Kuulkers1994,Church2012}. Their motion along the various branches is a function (although not monotonic) of the accretion rate (see \citealt{Mondal2018} and references). Their properties in terms of the QPOs observed and their branch-dependent spectral parameters are fairly well reported in literature (see \citealt{Mondal2018,Kuulkers1999,Piraino2001}). However very little is understood about the thermonuclear bursting behavior in these objects in their different Z positions. In fact, out of all the Z sources, only Cygnus X-2 and GX 17+2 exhibit Type-1 thermonuclear bursts, indicating unstable burning (see \citealt{Kuulkers2002} and references therein). This is in contrast with the other `Cyg-like' Z sources where stable burning at extremely high accretion rates inhibits occurrence of bursts. Moreover, burst studies in GX 17+2 using EXOSAT did not indicate any correlation between the bursts properties as a function of the position in the Z track \citep{Kuulkers2002}. \citet{Church2010} have assumed an Accretion Disc Corona (ADC) model to describe Cyg X-2 and the other sources from its class. Their model predicts that unstable thermonuclear burning is indeed possible during the FB.

During these AstroSat observations, Cyg X-2 has a source luminosity of 7.2$\times$10$^{37}$ erg/s. The horizontal profile of the HID is characteristic of a medium-to-high level accretion, indicative of either the FB or the HB. We seek out methods to distinguish between these two branches by conducting a spectral analysis followed by an examination of the power density spectra. The very first indication in favour of the FB comes from the fact that we observe X-ray bursts that are the result of unstable nuclear burning. \citet{Church2010} showed that in Cyg-like sources, the FB supports unstable nuclear burning. The non-burst light curve is also characterized by heavy dipping activity (Figure \ref{fig:h-ratio}) as also previously observed in \citet{Mondal2018}, \citet{DiSalvo2002} and \citet{kuulkers1996}. The presence of dips in the light curve is often observed during the FB \citep{Kuul1995, Mondal2018}. The results from non-burst persistent spectral fitting indicated that a thermal comptonization model along with a disk model was required to adequately describe the source. The inferred fluxes indicate luminosities close to 0.4$\times$L$_{edd}$. Such luminosities are achieved during the onset on an FB (similarly observed in \citealt{Mondal2018}). The luminosities derived from the fluxes in the compton component and the blackbody component are clearly indicative of the FB for Cyg X-2 and are consistent with those quoted in an RXTE analysis of this source (see Figure 4 from \citealt{Church2010}).  The power spectral density of the non-burst emission lacks the presence of QPOs. QPOs have been reported for Cyg X-2 largely during the NB and the HB with very few reports of FB QPOs as discussed in Section 3.2.3.

\citet{Church2010} show that during the soft apex (NB-early onset of FB in the CCD), the luminosities are not high enough to cause disruption of the inner disk allowing the inner disk to touch the neutron star surface. There will, therefore, be no oscillations in the edge that would give rise to kHz QPOs. Using Ginga data, \citet{Wijnands1997} study the X-ray variability in different intensity states and conclude that the QPOs are not detectable at the soft apex and in the FB. Overall, we conclude that these results indicate that Cyg X-2 was entering the early FB during the 2016 AstroSat observations.

A simple physical picture of the system can be visualized as follows. There is enough evidence from the dips observed in the X-ray light curves that Cyg X-2 is being viewed at a high inclination angle \citep{Vrtilek1988, Rykoff2010}. Such dipping sources are often also associated with a coronal component whose exact geometrical extent remains unclear to date \citep{Church1998, Church2010}. As the system  progresses into the FB with increased accretion rates, the inner disk is radiatively heated, gets puffed up and the corona expands \citep{Kuulkers1995,Wijnands1997}. The corona further condenses and cools and in the process forms clumps. This gives rise to X-ray absorption at certain orbital phases leading to dips in the X-ray light curve. With large accretion rates per unit area, and correspondingly higher X-ray fluxes, ignition conditions are achieved in faster succession and burst rates are higher.

Future observations of thermonuclear bursts from this class of Z sources like Cyg X-2 and GX 17+2 as they transition between their various Z positions would be beneficial in understanding their detonation properties as a function of the mass accretion rate.\\

\textit{Acknowledgements}:
The research is based on the results obtained from the AstroSat mission of the Indian Space Research Organisation (ISRO), archived at the Indian Space Science Data Centre (ISSDC). The authors thank Nandini from the RRI data center as well as the LAXPC poc team for assisting us with data procurement and reduction.

\bibliography{bibtex}{}
\bibliographystyle{elsarticle-harv}\biboptions{authoryear}

\appendix 

\section{Burst Oscillation search}

We present the results of search for Burst Oscillations in the 6 bursts detected using LAXPC. These dynamic power spectra were generated using the GHATS timing package\footnote{http://www.brera.inaf.it/utenti/belloni/GHATS\_{Package}/Home.html}. Details of the processing are given in Section 3.1.1. 

\begin{figure}[ht]
	\centering
	\includegraphics[scale=0.16,angle=0,trim={0cm 1cm 1cm 1cm},clip=true]{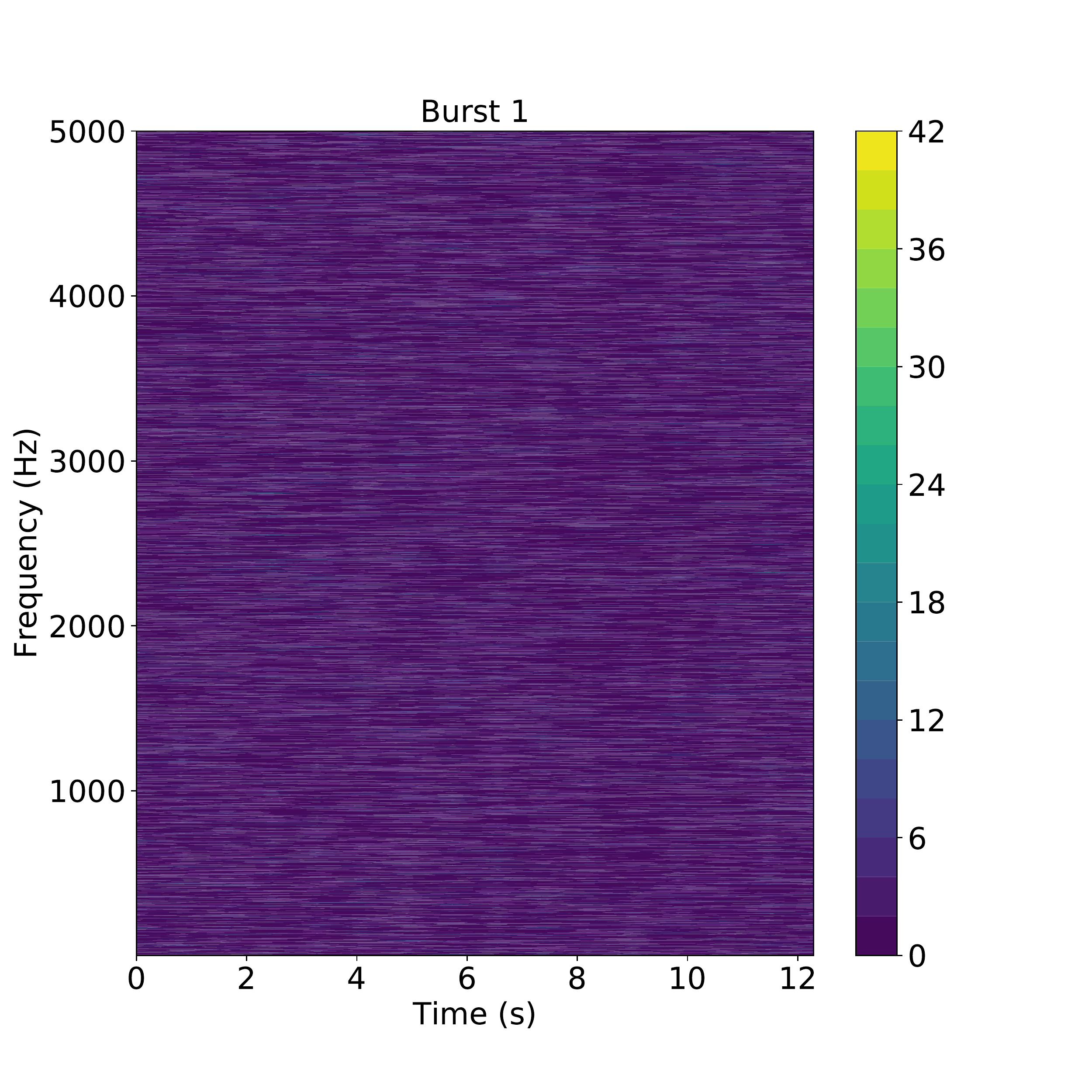}
	\includegraphics[scale=0.16,angle=0,trim={0cm 1cm 1cm 1cm},clip=true]{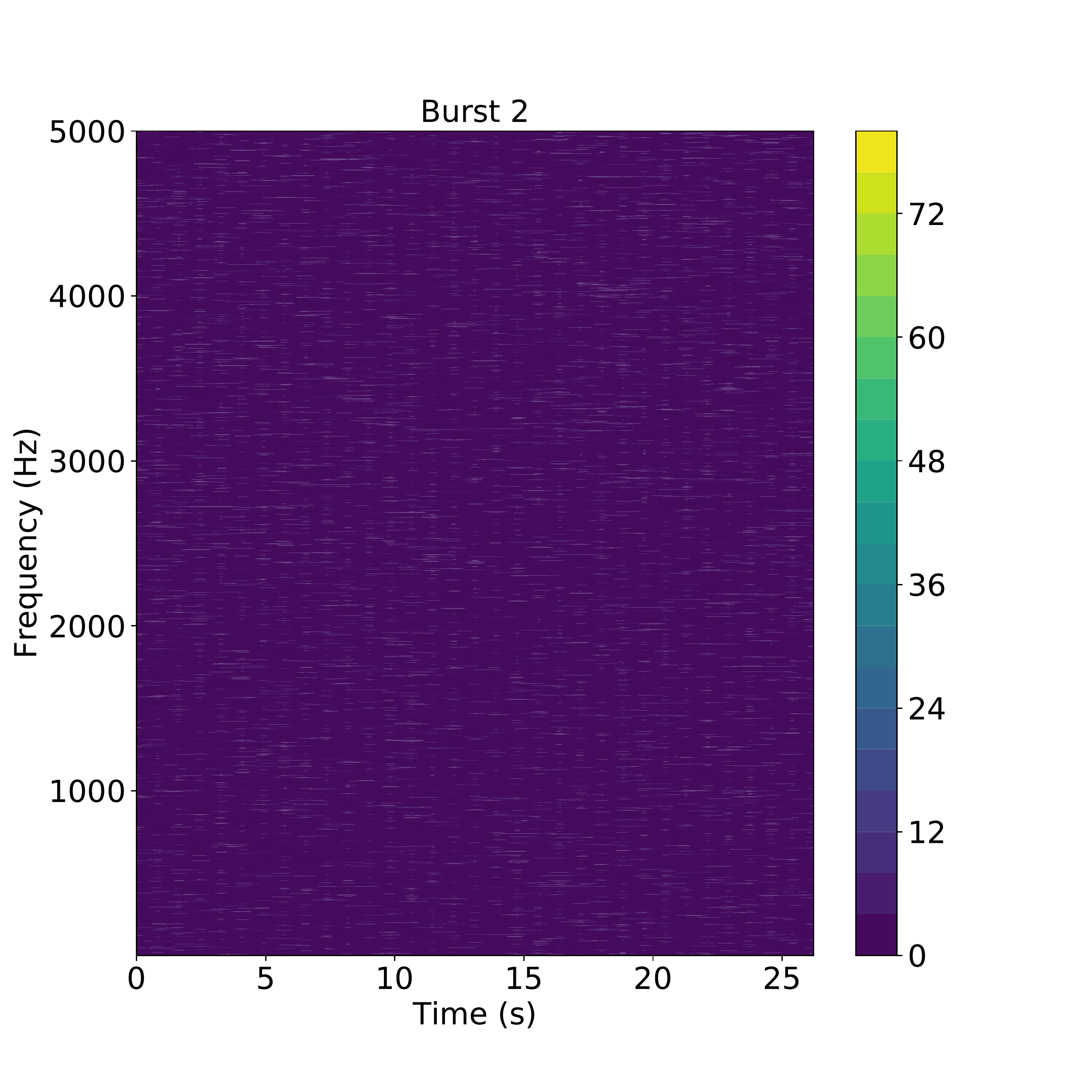}
	
	\includegraphics[scale=0.16,angle=0,trim={0cm 1cm 1cm 1cm},clip=true]{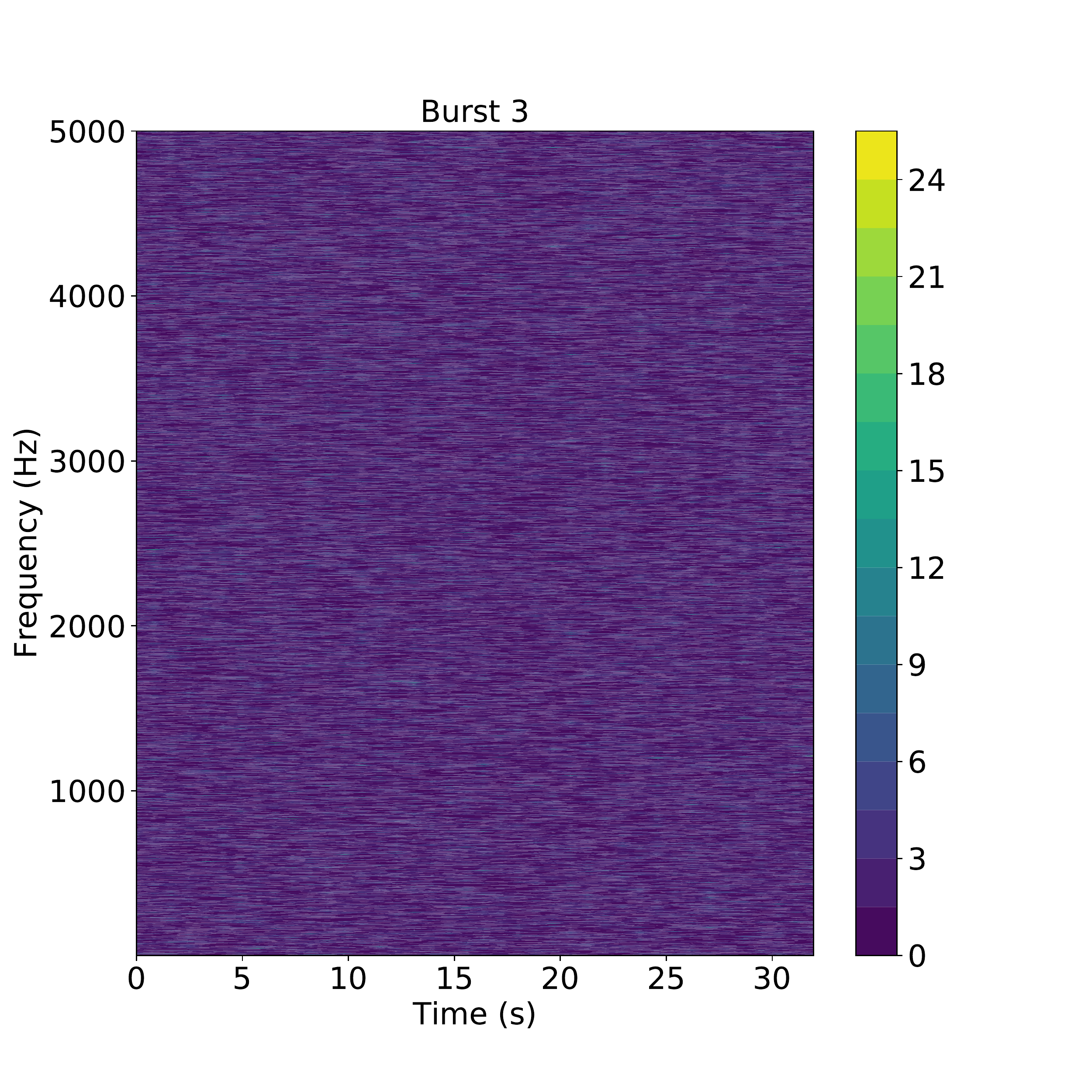}
	\includegraphics[scale=0.16,angle=0,trim={0cm 1cm 1cm 1cm},clip=true]{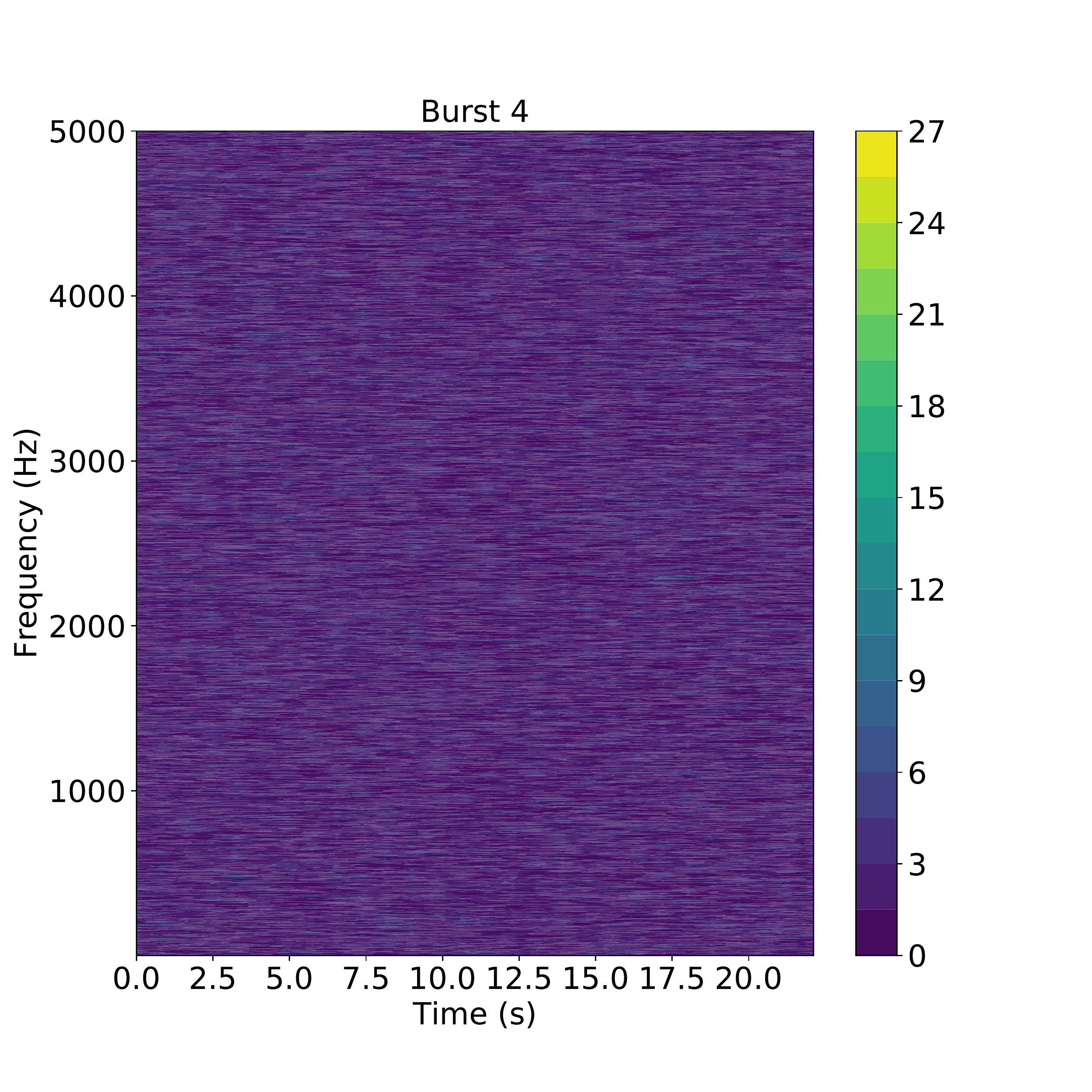}
	
	\includegraphics[scale=0.16,angle=0,trim={0cm 1cm 1cm 1cm},clip=true]{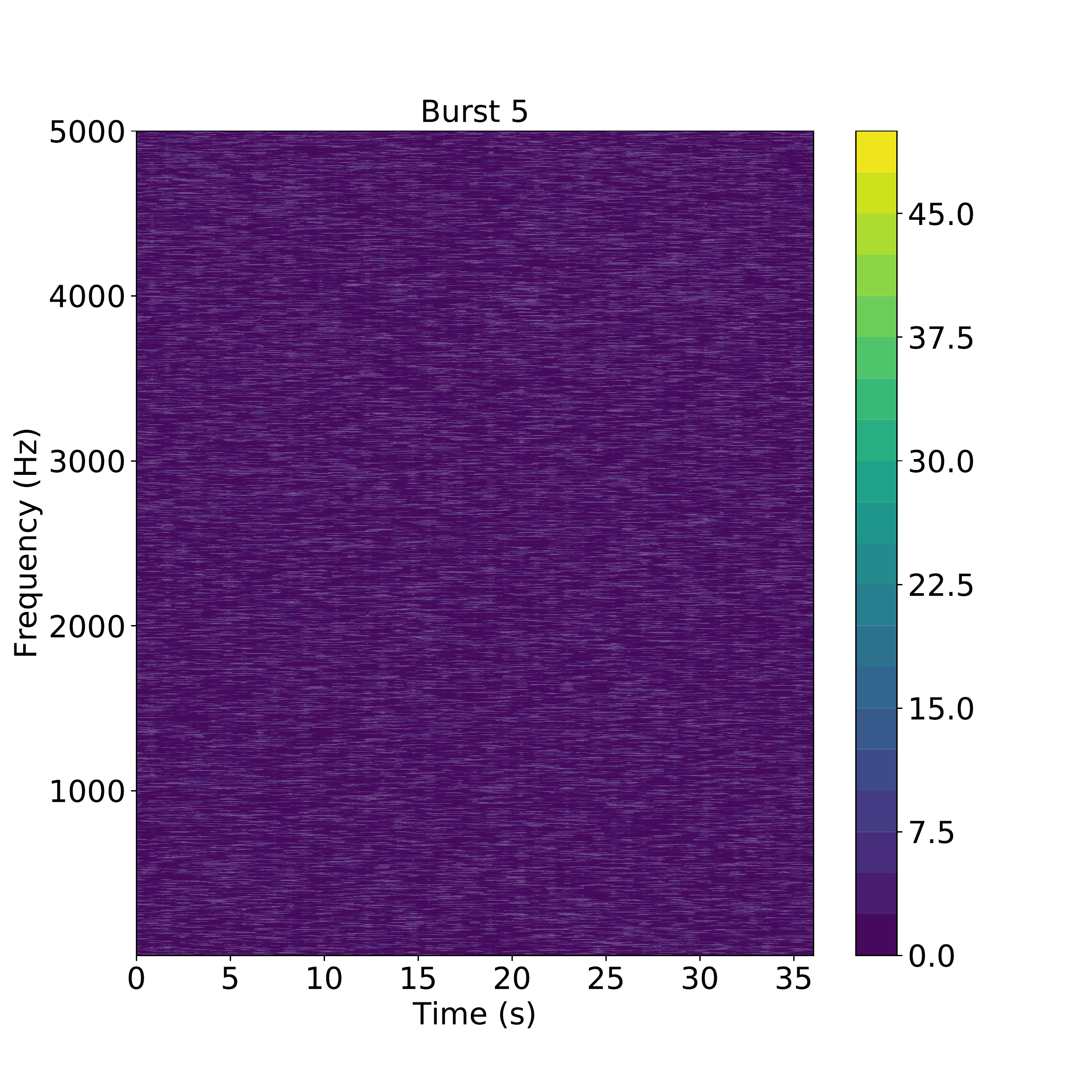}
	\includegraphics[scale=0.16,angle=0,trim={0cm 1cm 1cm 1cm},clip=true]{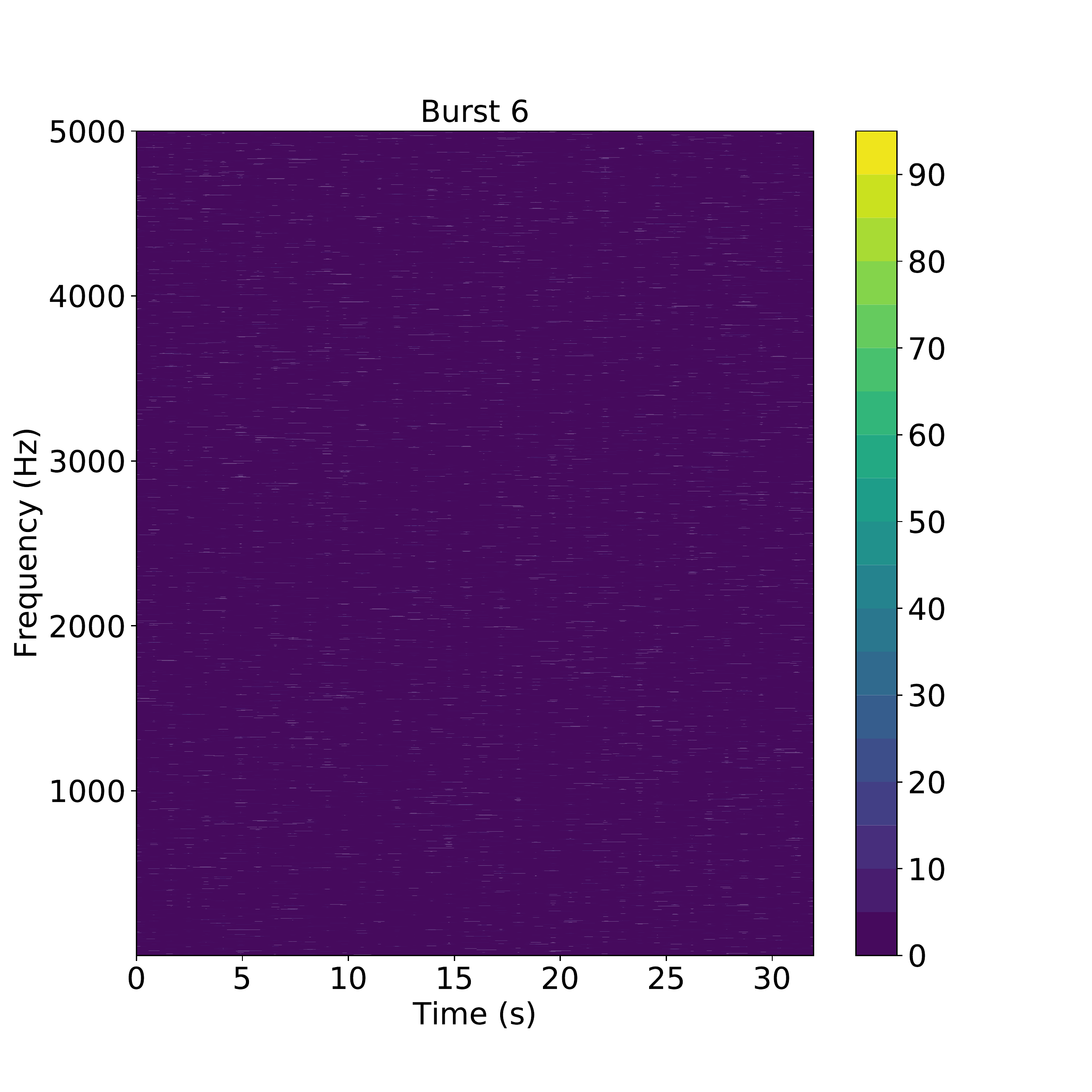}
	
	\caption{Dynamic power spectra constructed for all the six type-1 X-ray bursts detected from Cyg X-2 using the GHATS X-ray timing software. No signal is obtained between 1- 5 kHz. }
	\label{fig:dynps}
\end{figure}

\end{document}